\documentclass[acmtog]{acmart}

\usepackage{kotex}
\usepackage{comment}
\usepackage{color, colortbl}
\usepackage{makecell}
\usepackage{soul}

\usepackage{amsmath}
\usepackage{algpseudocode}
\usepackage{algorithm}
\algnewcommand{\LeftComment}[1]{\Statex \(\triangleright\) #1}

\usepackage{booktabs}
\usepackage{multirow}
\usepackage{graphicx}
\usepackage{balance}

\newcommand{\Eq}[1]  {Eq.\ (\ref{equ:#1})}

\newcommand{\Fig}[1] {Figure \ref{fig:#1}}

\newcommand{\Tbl}[1]  {Table \ref{tbl:#1}}

\newcommand{\Sec}[1] {Section \ref{sec:#1}}

\newcommand{\Etal}   {et al.}

\setcopyright{acmlicensed}\acmJournal{TOG}
\acmYear{2022}\acmVolume{41}\acmNumber{6}\acmArticle{216}\acmMonth{12} \acmDOI{10.1145/3550454.3555511}

\begin{document}

\title[LaplacianFusion]{LaplacianFusion: Detailed 3D Clothed-Human Body Reconstruction}

\author{Hyomin Kim}\orcid{0000-0002-2162-4627}
\email{min00001@postech.ac.kr}
\author{Hyeonseo Nam}\orcid{0000-0003-4033-901X}
\email{hyeonseo.nam@postech.ac.kr}
\author{Jungeon Kim}\orcid{0000-0003-4212-1970}
\email{jungeonkim@postech.ac.kr}
\author{Jaesik Park}\orcid{0000-0001-5541-409X}
\email{jaesik.park@postech.ac.kr}
\author{Seungyong Lee}\orcid{0000-0002-8159-4271}
\email{leesy@postech.ac.kr}
\affiliation{\institution{POSTECH}\city{Pohang}\country{Korea}}
\renewcommand{\shortauthors}{H. Kim, et al.}

\begin{abstract}
We propose \textit{LaplacianFusion}, a novel approach that reconstructs detailed and controllable 3D clothed-human body shapes from an input depth or 3D point cloud sequence.
The key idea of our approach is to use Laplacian coordinates, well-known differential coordinates that have been used for mesh editing, for representing the local structures contained in the input scans, instead of implicit 3D functions or vertex displacements used previously.
Our approach reconstructs a controllable base mesh using SMPL, and learns a surface function that predicts Laplacian coordinates representing surface details on the base mesh.
For a given pose, we first build and subdivide a base mesh, which is a deformed SMPL template, and then estimate Laplacian coordinates for the mesh vertices using the surface function.
The final reconstruction for the pose is obtained by integrating the estimated Laplacian coordinates as a whole.
Experimental results show that our approach based on Laplacian coordinates successfully reconstructs more visually pleasing shape details than previous methods. The approach also enables various surface detail manipulations, such as detail transfer and enhancement.
\end{abstract}



\begin{teaserfigure}
  \includegraphics[width=\textwidth]{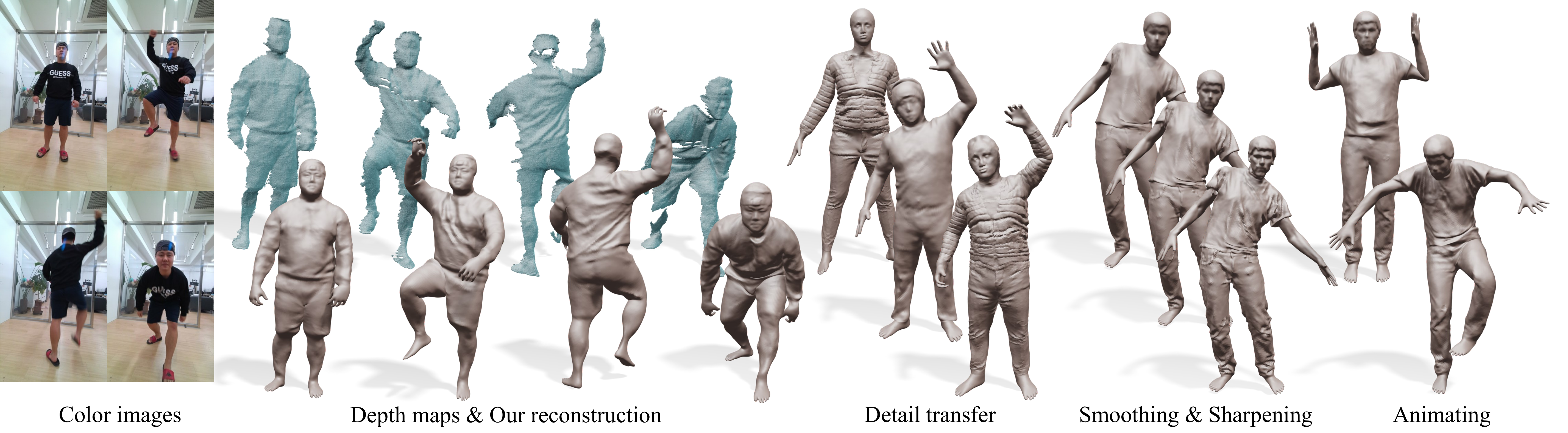}
  \caption{LaplacianFusion reconstructs a detailed and controllable 3D clothed-human model from a point cloud sequence that contains challenging dynamic motions. 
  The reconstructed model consists of a base mesh that fits the input scans and a surface function that predicts Laplacian coordinates representing the details on the body surface.
  The model can be animated by controlling the human pose applied to the base mesh, and the surface function enables various shape manipulations, such as detail transfer, smoothing, and sharpening, in addition to detail reconstruction.
  }
  \label{fig:teaser}
\end{teaserfigure}

\maketitle

\section{Introduction}
\label{sec:Intro}
3D reconstruction of clothed human models is crucial for reproducing digital twins of real world to give the user a sense of reality and immersion.
Clothed human models are useful for various applications, such as entertainment, virtual reality, and the movie industry.
In particular, with the surging demands for social connections in virtual spaces, it is valuable to produce realistic 3D human models in a typical capturing setup.

Parametric human models have been proposed to reconstruct 3D full body shapes for different poses.
Among them, SMPL~\cite{loper2015smpl} is a representative model and represents a human model with shape and pose parameters that are applied to a single template mesh with fixed topology.
In SMPL, shape deformations are obtained by linear blend skinning of the template mesh and cannot be detailed enough for depicting surface details of clothed human models.
This limitation also applies to other parametric models of human shapes~\cite{frankmodel, SMPL-X:2019}.

Recent learning-based methods for clothed human reconstruction utilize implicit 3D functions~\cite{saito2021scanimate, wang2021metaavatar}, but they learn a function defined in a 3D space and need an additional polygon extraction step to provide a 3D mesh as the output.
An explicit point cloud representation has been used to reconstruct loose clothes~\cite{POP:ICCV:2021,ma2021SCALE}, but this approach also needs the surface reconstruction step to produce a 3D mesh that can be directly used for applications.
On the other hand, Burov \Etal~\shortcite{burov2021dsfn} use an explicit mesh representation and train a Dynamic Surface Function Network (DSFN) to estimate vertex displacements for a template mesh. 
However, DSFN may not fully exploit the surface details in the input scans due to the spatial regularization constraint needed for training.

In this paper, we present {\em LaplacianFusion}, a novel framework that reconstructs a {\em detailed} and {\em controllable} 3D clothed human model from an input depth or point cloud sequence.
Our key idea is to use differential coordinates, instead of implicit 3D function or vertex displacements, for representing local structures of surface details. 
For the differential coordinates, we use Laplacian coordinates~\cite{karni2000spectral, alexa2003differential, sorkine2003high} that have been widely applied to mesh processing and editing~\cite{lipman2004differential, sorkine2004laplacian}.
Intuitively, Laplacian coordinates are the difference between a point and the average of its neighbors, so they can naturally encode local shape variations.

In our LaplacianFusion framework, the reconstructed human model is expressed as combination of a controllable base mesh and a surface function using a multi-layer perceptron (MLP).
In the training phase, we first reconstruct a base mesh sequence based on SMPL that fits the input scans.
We then train an MLP function by fusing the Laplacian values estimated at input points on the surface of the SMPL template mesh.
As a result, the MLP function learns to predict Laplacian coordinates representing the details on the body surface. 
To reconstruct a detailed human model for a given pose, we start from a SMPL template mesh in the canonical space and obtain a base mesh by deforming the SMPL template using the pose parameters. 
We then subdivide the base mesh to have enough vertices and estimate Laplacian coordinates for the vertices using the learned MLP function.
Finally, the detailed output mesh is obtained by globally integrating the estimated Laplacian coordinates as a whole.
In this paper, we call the MLP function {\em neural surface Laplacian function}.

We aim for a natural capture scenario where the subject freely performs actions during the capture.
Our approach can handle both full and partial views of point clouds that are captured by a dome-shaped multi-camera setup~\cite{fvv} and a single RGB-D camera~\cite{KinectAzure}, respectively. 
The reconstructed 3D models are controllable as the base mesh and the neural surface Laplacian function are conditioned on SMPL pose parameters.
Our approach restores the surface details of a clothed human body better than the recent explicit surface-based approach, DSFN~\cite{burov2021dsfn}.
In addition, due to the differential representation and a fixed-topology base mesh, our approach can be easily adapted for other applications, including detail transfer, detail enhancement, and texture transfer.
Our codes are publicly available.\footnote{\href{https://github.com/T2Kim/LaplacianFusion}{https://github.com/T2Kim/LaplacianFusion}}

Our main contributions can be summarized as follows:
\begin{itemize}
\item We propose \textit{LaplacianFusion}, a novel framework for reconstructing surface details of a clothed human body model. Our framework can handle both partial and full-body point cloud sequences.
\item We introduce an approach to learn Laplacian coordinates representing surface details from scanned points using a MLP.
\item Our reconstructed model is controllable by pose parameters and supports various shape manipulations, including detail transfer.
\end{itemize}

\section{Related Work}
\subsection{Parametric models for 3D human shapes and clothes}
\label{param_model}
\paragraph{Human body models}
PCA-based parametric models have been proposed for handling human body and pose variations: SMPL~\cite{loper2015smpl, MANO:SIGGRAPHASIA:2017, SMPL-X:2019}, GHUM~\cite{xu2020ghum}, and Frank model~\cite{frankmodel}. 
These models can handle body shape variations and pose-dependent shape deformations that cannot be modeled with Linear Blend Skinning (LBS)~\cite{lewis2000pose}. The models are suitable for expressing human shapes with coarse meshes, but they alone are not enough for containing rich details.

\paragraph{Clothed human models}
\label{cloparam}
Several approaches represented clothed humans by extending parametric human models. 
SMPL \cite{loper2015smpl} has been extended to express clothed deformations by directly adding a displacement vector to each vertex~\cite{alldieck2019learning, alldieck2018detailed, alldieck2018video, bhatnagar2020ipnet}.
CAPE~\cite{ma2020cape} proposed an extended model by adding a cloth style term to SMPL.
Other approaches~\cite{de2010stable, guan2012drape, pons2017clothcap, bhatnagar2019multi, tiwari20sizer, xiang2020monoclothcap} used additional parametric models for representing clothes on top of a parametric human model. 
Additional approaches for expressing surface details include GAN-based normal map generation~\cite{lahner2018deepwrinkles}, RNN-based regression~\cite{santesteban2019learning}, and style-shape specific MLP functions~\cite{patel2020tailornet}. However, these approaches are limited to several pre-defined clothes and cannot recover detailed human shapes with arbitrary clothes from input scans.

\subsection{Implicit clothed human reconstruction}
\paragraph{Volumetric implicit representations}
Truncated signed distance function (TSDF) is a classical implicit representation for reconstruction. TSDF-based approaches that warp and fuse the input depth sequence onto the canonical volume have been proposed for recovering dynamic objects~\cite{dynamicfusion, volumedeform}. 
This volume fusion mechanism is extended to be conditioned with the human body prior for representing a clothed human~\cite{BodyFusion,yu2018DoubleFusion}. 
Optimized approaches for real-time performance capture have also been proposed~\cite{dou2016fusion4d, dou2017motion2fusion, habermann2020deepcap, yu2021function4d}.

\paragraph{Neural implicit representations}
MLP-based neural implicit representation has been actively investigated for object reconstruction~\cite{Park_2019_CVPR, Mescheder2019occnet, Chen2019LearningIF}. PIFu \cite{saito2019pifu, saito2020pifuhd} firstly adopted this representation for reconstructing static clothed human from a single image. DoubleField~\cite{Shao_2022_CVPR} uses multi-view RGB cameras and improves visual quality by sharing the learning space for geometry and texture. With depth image or point cloud input, multi-scale features~\cite{chibane20ifnet, bhatnagar2020ipnet} and human part classifiers~\cite{bhatnagar2020ipnet} have been used for reconstructing 3D human shapes. Li et al.~\shortcite{li2021posefusion} proposed implicit surface fusion from a depth stream and enabled detailed reconstruction even for invisible regions.

Neural parametric models have also been proposed for modeling shape and pose deformations. NASA~\cite{deng2019NASA} learns pose-dependent deformations using part-separate implicit functions. LEAP~\cite{mihajlovic2021LEAP} and imGHUM~\cite{alldieck2021imghum} learn parametric models that can recover shape and pose parameters for SMPL~\cite{loper2015smpl} and GHUM~\cite{xu2020ghum}, respectively. NPMs~\cite{palafox2021npms} encodes shape and pose variations into two disentangled latent spaces using auto-decoders~\cite{Park_2019_CVPR}. SPAMs~\cite{Palafox_2022_CVPR} introduces a part-based disentangled representation of the latent space.

For subject-specific clothed human reconstruction from scans, recent implicit methods define shape details in a canonical shape and use linear blend skinning to achieve both detail-preservation and controllability. Neural-GIF~\cite{tiwari2021neural} learns a backward mapping network for mapping points to the canonical space and a displacement network working in the canonical space.
In contrast, SNARF~\cite{Chen_2021_ICCV} proposed a forward skinning network model to better handle unseen poses. 
SCANimate~\cite{saito2021scanimate} learns forward and backward skinning networks with a cycle loss to reconstruct disentangled surface shape and pose-dependent deformations. 
MetaAvatar~\cite{wang2021metaavatar} proposed an efficient pipeline for subject-specific fine-tuning using meta-learning. 

These implicit representations are topology-free and can handle vastly changing shapes, such as loose clothes. However, they individually perform shape reconstruction at every frame and cannot provide temporally consistent mesh topology needed for animation.
In addition, this approach needs dense point sampling in a 3D volume to train implicit functions and is computationally heavy.
    
\begin{table}[t]
\centering
\caption{Comparison of approaches that reconstruct 3D clothed-human shapes. Our approach explicitly handles a rigged 3D mesh model, so the reconstruction is animatable and texture map can be readily applied.}
\vspace{-3.3mm}
\resizebox{.995\linewidth}{!}{
\begin{tabular}{c|c|c|c|c|c|c}
& {Method} & \makecell[c]{Shape\\representation} & \rotatebox[origin=c]{90}{2.5D input} & \rotatebox[origin=c]{90}{3D input} & \rotatebox[origin=c]{90}{\makecell[c]{Animation\\ready}} &  \rotatebox[origin=c]{90}{\makecell[c]{Texture map\\ready}} \\

\hline
\multirow{12}{*}{\rotatebox[origin=c]{90}{Implicit}} & DynamicFusion \shortcite{dynamicfusion} & SDF  & \checkmark & & &  \\
 & BodyFusion \shortcite{BodyFusion} & SDF & \checkmark & & & \\
 & NASA \shortcite{deng2019NASA} & Occupancy & & \checkmark & \checkmark \\
 & IF-Net \shortcite{chibane20ifnet} & Occupancy & \checkmark & \checkmark & &  \\
 & IP-Net \shortcite{bhatnagar2020ipnet} & Occupancy & \checkmark & \checkmark & \checkmark &\\
 & NPMs \shortcite{palafox2021npms} & SDF & \checkmark & \checkmark & \checkmark & \\
 & Neural-GIF \shortcite{tiwari2021neural} & SDF & & \checkmark & \checkmark & \\
 & SCAnimate \shortcite{saito2021scanimate} & SDF & & \checkmark & \checkmark & \\
 & MetaAvatar \shortcite{wang2021metaavatar} & SDF & \checkmark & \checkmark & \checkmark & \\
 & SNARP \shortcite{Chen_2021_ICCV} & SDF & & \checkmark & \checkmark & \\
 & POSEFusion \shortcite{li2021posefusion} & Occupancy & \checkmark & & \checkmark & \\
 & LEAP \shortcite{mihajlovic2021LEAP} & Occupancy & & \checkmark & \checkmark & \\
\hline
\multirow{5}{*}{\rotatebox[origin=c]{90}{Explicit}}                 
 & CAPE \shortcite{ma2020cape} & Coord. (vertex) & & \checkmark & \checkmark & \checkmark\\
 & SCALE \shortcite{ma2021SCALE} & Coord. (patch) & & \checkmark & \checkmark & \\
 & PoP \shortcite{POP:ICCV:2021} & Coord. (point) & & \checkmark & \checkmark & \\
 & DSFN \shortcite{burov2021dsfn} & Coord. (vertex) & \checkmark & & \checkmark & \checkmark\\
\rowcolor{yellow}
 & Ours & Coord. + Laplacian & \checkmark & \checkmark & \checkmark & \checkmark \\
\end{tabular}
}
\vspace{-3mm}
\label{tbl:related work}
\end{table}

\subsection{Explicit clothed human reconstruction.}
Explicit representations for reconstructing clothed humans have been developed mainly for handling geometric details on the template mesh of a parametric model such as SMPL ~\cite{loper2015smpl}. 
\paragraph{Point-based} 
SCALE~\cite{ma2021SCALE} recovers controllable clothed human shapes by representing local details with a set of points sampled on surface patches. To avoid artifacts of the patch-based approach at patch boundaries, PoP~\cite{POP:ICCV:2021} represents local details using point samples on a global 2D map. Point cloud representation has a flexible topology and can cover more geometric details. However, this representation does not provide an explicit output mesh.
\paragraph{Mesh-based} 
For subject-specific human reconstruction with a depth sequence, DSFN~\cite{burov2021dsfn} represents surface details using vertex offsets on a finer resolution mesh obtained by subdividing the SMPL template mesh. This approach is the closest to ours. The main difference is that our method represents surface details using Laplacian coordinates, instead of vertex offsets. Experimental results show that Laplacian coordinates are more effective for recovering surface details than vertex offsets (\Sec{exp}).

\Tbl{related work} compares our method with related ones in terms of possible inputs and desirable properties. 
\section{Preliminary}
\label{sec:background}

\begin{figure}[t]
	\centering
	\includegraphics[width=0.45\textwidth]{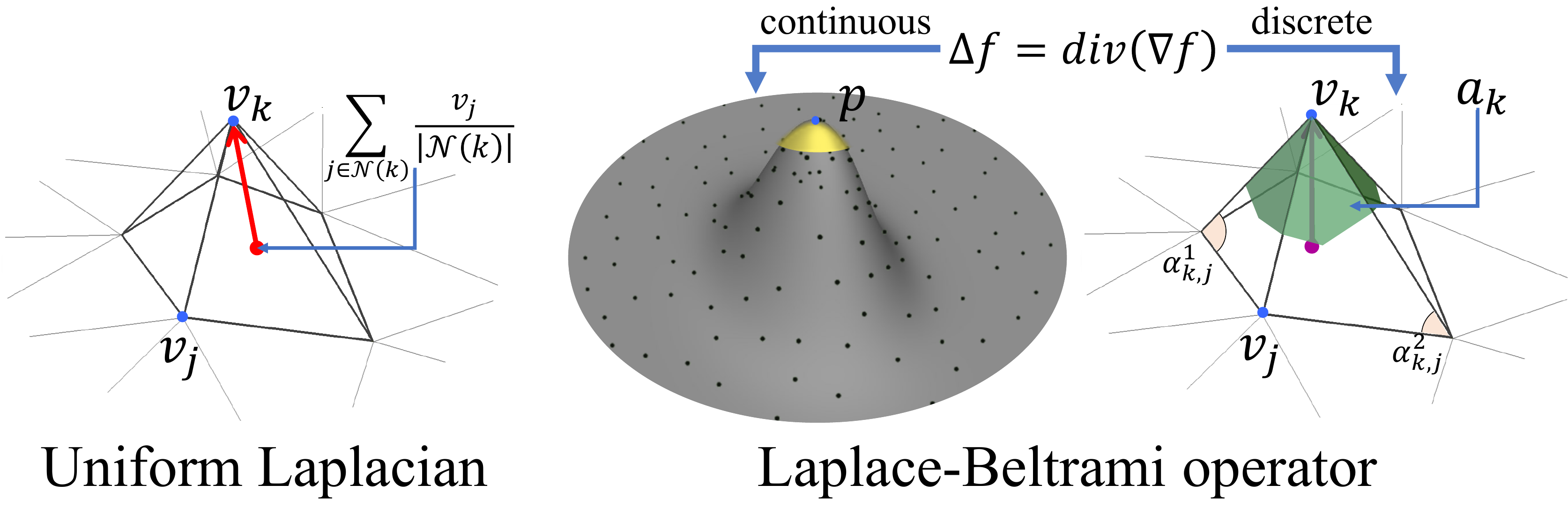}\\
	\caption{Computing Laplacian coordinates on a mesh and a point cloud. (left) Mesh editing methods usually calculate Laplacian coordinates (red arrow) using uniform average of neighbor vertices on a mesh. (middle) Our input is a point cloud, and we approximate Laplacian coordinates by fitting a quadratic polynomial to a small region (yellow region) in the input point cloud and then using Laplace-Beltrami operator that produces differentials of a smooth real function $f$. (right) The approximated Laplacian coordinates differ from uniform Laplacian coordinates, so we utilize the discrete Laplace-Beltrami operator to convert Laplacian coordinates to absolute vertex positions.
	}
	\label{fig:explanation[laplace]}
\end{figure}

\begin{figure*}[t]
		\includegraphics[width=1\textwidth]{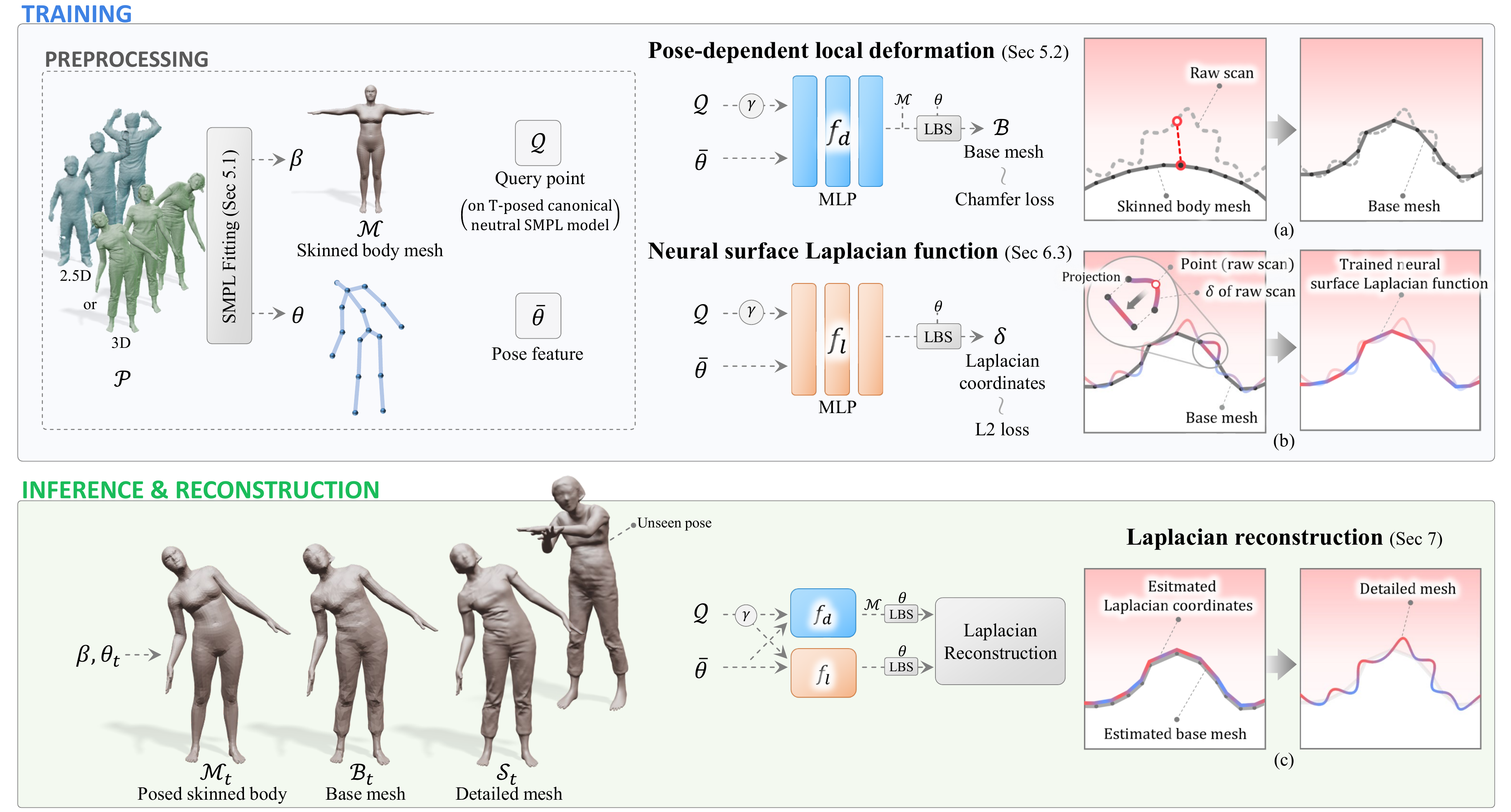}\\
\caption{System overview. (top blue box) LaplacianFusion takes a 2.5D depth sequence (mint color) or full-body 3D point clouds (olive color), and produces detailed meshes. In the training phase, (top left) we initially align the SMPL mesh to the input point clouds and obtain the skinned body meshes. (a) Then, we learn pose-dependent local deformations for the vertices of the skinned body meshes to accurately fit the input data. (b) To capture surface details, we set training pairs by projecting each raw scan to the base mesh, then learn neural surface Laplacian function that predicts pose-dependent Laplacian coordinates on the surface of a base mesh. (bottom green box) In the reconstruction phase, we can recover and animate the final 3D shapes using the base mesh controlled by pose parameters and the neural surface Laplacian function estimating surface details. (c) We conduct Laplacian reconstruction to convert the estimated Laplacian coordinates to vertex positions. Note that the red and blue colors illustrated on the line segments in (b) and (c) represent Laplacian coordinates. 
}
\label{fig:overallProcess}
\end{figure*}

\subsection{Laplacian coordinates from a mesh}
\label{sec:LapMesh_editing}
In the graphics literature, recovering an unknown function from differential quantities (Laplacian) has become widely known through Poisson image editing~\cite{perez2003poisson}.
This technique was successfully expanded to the 3D mesh domain, especially for mesh editing~\cite{lipman2004differential, sorkine2004laplacian}. In mesh editing, the differential quantities are used to encode vertex coordinates and called {\em Laplacian coordinates}. Mesh editing based on Laplacian coordinates includes three steps: Encoding Laplacian coordinates from the original mesh, interactive editing of control points, and converting Laplacian coordinates into absolute vertex positions of the target mesh while satisfying the positional constraints imposed by edited control points. In the following, we briefly introduce the encoding and converting steps.

Let the original mesh $\mathcal{M}=\{\mathcal{V}, \mathcal{F}\}$ be described by the vertex set $\mathcal{V}$ and the triangle set $\mathcal{F}$, where $\mathcal{V}=\{\mathbf{v}_k\mid k = 1, \dots, K\}$. $\mathbf{v}_k$ denotes the position of the $k$-th vertex and $K$ is the number of vertices. Uniform Laplacian coordinates $\boldsymbol{\widehat{\delta}}_k\in\mathbb{R}^3$ are defined by:
\begin{equation}
\begin{split}
\label{equ:LapCoord_meshedit}
    \boldsymbol{\widehat{\delta}}_k=\sum_{j\in \mathcal{N}(k)}\widehat{w}_k{\left(\mathbf{v}_k-\mathbf{v}_j\right)},
\end{split}
\end{equation}
where $\widehat{w}_k = \frac{1}{|\mathcal{N}(k)|}$ indicates uniform weights, and $\mathcal{N}(k)$ denotes the set of adjacent vertices of the $k$-th vertex (\Fig{explanation[laplace]} left). Regarding all vertices, this equation can be represented in a matrix form: $[\boldsymbol{\widehat{\delta}}_1, \dots,\boldsymbol{\widehat{\delta}}_K]^T=\widehat{\mathbf{L}}[\mathbf{v}_1, \dots,\mathbf{v}_K]^T$, where $\widehat{\mathbf{L}}$ is the uniform Laplacian matrix.
Notably, the matrix $\widehat{\mathbf{L}}$ has rank $K-1$, so $\{\boldsymbol{\widehat{\delta}}_k\}$ can be converted into $\mathcal{V}$ by taking the specified position of a selected vertex as the boundary condition and solving a linear system.
For example, when fixing the $i$-th vertex, we can form a sparse linear system $\mathbf{A}\mathbf{x}=\mathbf{b}$, where $\mathbf{A}=[\widehat{\mathbf{L}}\,^T, \mathbf{1}_i]^T$ and $\mathbf{b}=[\boldsymbol{\widehat{\delta}}_1, \dots,\boldsymbol{\widehat{\delta}}_K, \mathbf{v}_i]^T$. $\mathbf{1}_i$ denotes one-hot encoding, where the $i$-th element is one.

\subsection{Laplacian coordinates from a point cloud}
\label{sec:LapPCD}
In this paper, we compute Laplacian coordinates from raw 3D scan data and use them for shape detail reconstruction.
Then, we need an alternative approach to \Eq{LapCoord_meshedit} for computing Laplacian coordinates as a point cloud does not have edge connectivity. We may consider directly building edges from the point set, but it may generate a noisy and non-manifold mesh. To resolve this difficulty, Liang et al.~\shortcite{liang2012geometric} defined the Laplace-Beltrami operator on a point cloud by fitting a quadratic function for the local neighborhood of a point and computing the differentials of the function (\Fig{explanation[laplace]} middle). However, the Laplace-Beltrami operator computes Laplacian coordinates using a {\em continuous} function that reflects the {\em non-uniform} local shape in the neighborhood, which differs from the {\em discrete uniform} Laplacian coordinates in \Eq{LapCoord_meshedit}. 
Therefore, Laplacian coordinates calculated by the Laplace-Beltrami operator need to be converted into mesh's vertex positions differently.

Let's assume that we have a mesh $\mathcal{M}=\{\mathcal{V}, \mathcal{F}\}$. We can calculate the discrete Laplace-Beltrami operator on the mesh (\Fig{explanation[laplace]} right) as follows~\cite{meyer2003discrete}:
\begin{equation}
\begin{split}
\label{equ:cotLap}
    \boldsymbol{\delta}_k=\Delta_{\mathcal{M}}(\mathbf{v}_k)&=\frac{1}{a_k}
    \sum_{j\in \mathcal{N}(k)}{\frac{cot(\alpha^1_{k,j})+cot(\alpha^2_{k,j})}{2}\left(\mathbf{v}_k-\mathbf{v}_j\right)},
\end{split}
\end{equation}
where $\boldsymbol{\delta}_k$ is non-uniform Laplacian coordinates, $\Delta_{\mathcal{M}}$ is the discrete Laplace-Beltrami operator, $cot(\cdot)$ denotes cotangent function, $a_k$ is the Voronoi area of $\mathbf{v}_k$, and $\alpha^1_{k,j}$ and $\alpha^2_{k,j}$ are the two angles opposite to the edge $\{k,j\}$ on $\mathcal{M}$. 
Similarly to \Eq{LapCoord_meshedit}, \Eq{cotLap} can be represented in a matrix form 
$[\boldsymbol{\delta}_1, \dots,\boldsymbol{\delta}_K]^T=\mathbf{L}[\mathbf{v}_1, \dots,\mathbf{v}_K]^T$, where $\mathbf{L}$ is the non-uniform Laplacian matrix, and we can convert $\{\boldsymbol{\delta}_k\}$ into $\mathcal{V}$ by solving a linear system with a boundary condition. 

Compared with global coordinates, which represent exact spatial locations, Laplacian coordinates naturally encode local shape information, such as the sizes and orientations of local details~\cite{sorkine2006differential}. In mesh editing, this property has been used to retain desirable local shapes during the editing process by encoding Laplacian coordinates from the original mesh and constraining the edited mesh to follow the encoded Laplacian coordinates.
In this paper, we use the property for shape reconstruction rather than editing. We extract and encode the local details from the input scan data using Laplacian coordinates and restore the desirable details on the subdivided base mesh by integrating the encoded Laplacian coordinates.
\section{Overview}
\label{sec:overall_process}
\Fig{overallProcess} shows the overall process of our LaplacianFusion framework.
The input is a sequence of point clouds $\{\mathcal{P}_t\}_{t=\{1,...,T\}}$, each of which contains either 2.5D depth map or 3D full scan of a clothed human body with motion.
Our pipeline starts from the SMPL template mesh in the canonical T-pose (\Sec{basemesh}). SMPL provides the shape prior for the reconstructed models and enables the controllability using pose parameters. Given an input sequence, we estimate a single SMPL shape parameter $\boldsymbol{\beta}$ and per-frame SMPL pose parameters $\boldsymbol\theta_t$, and obtain the posed skin body $\mathcal{M}_t$ for each frame $t$.
Since the mesh $\mathcal{M}_t$ may not accurately fit the input point cloud, we add pose-dependent local deformations to improve the fitting and obtain the base mesh $\mathcal{B}_t$ (\Sec{Method[PBBM]}), where the local deformations are estimated using an MLP function $f_d$ that computes vertex displacements for $\mathcal{M}_t$.
On top of the base mesh $\mathcal{B}_t$, our novel {\em neural surface Laplacian function} $f_l$ predicts Laplacian coordinates that encode the surface details of a clothed human model (\Sec{LapSurfFunc}).
Finally, we reconstruct the detailed output mesh $\mathcal{S}_t$ by integrating Laplacian coordinates estimated at the vertices of the subdivided base mesh (\Sec{Method[Recon]}). 
Note that $f_d$ and $f_l$ are functions of both 3D positions and pose parameters as the local deformations and cloth details such as wrinkles are pose-dependent.
We outline the training and inference phases for the two surface functions below.

\paragraph{Training phase}
We sequentially train the two surface functions because 3D points on a base mesh $\mathcal{B}$ obtained by $f_d$ are used as the input of $f_l$. We train $f_d$ to learn a deformation field that minimizes the Chamfer distance~\cite{barrow1977parametric} between the input point cloud and the base mesh $\mathcal{B}$. 
To train $f_l$, for each point in the input point cloud, we calculate approximate Laplacian coordinates as described in \Sec{LapPCD}, and find the corresponding point on the base mesh $\mathcal{B}$ to assign the calculated Laplacian coordinates.
$f_l$ is then trained to learn the assigned Laplacian coordinates on $\mathcal{B}$ by mininmizing the L2 loss.
 
\paragraph{Inference phase} 
We can infer the surface details for a particular pose using the learned MLP functions $f_d$ and $f_l$.
We first obtain the posed skinned body $\mathcal{M}$ by applying the given pose parameter to the SMPL template mesh.
We then apply the local deformation function $f_d$ to $\mathcal{M}$ and obtain the pose-dependent base mesh $\mathcal{B}$. Since the vertex number of the SMPL template is insufficient to represent surface details of a clothed human, we subdivide the base mesh $\mathcal{B}$.
We estimate Laplacian coordinates for each vertex of the subdivided $\mathcal{B}$ using neural surface Laplacian function $f_l$. 
Finally, we reconstruct a detailed clothed human model by integrating the estimated Laplacian coordinates.
Note that the pose parameter used in the inference phase can be arbitrary, and it does not have to be one of the pose parameters estimated from the input frames (\Sec{animate}).

\section{Pose-dependent base mesh}

\subsection{Skinned body acquisition}
\label{sec:basemesh}
Our approach starts with building a skinned body.
We adopt the SMPL model~\cite{loper2015smpl} that can readily manipulate 3D human shape using identity-dependent parameters $\boldsymbol{\beta}$ and pose-dependent parameters $\boldsymbol{\theta}$.
SMPL supports rigging and skinning, and the template mesh can be deformed with an arbitrary pose.
As in the SMPL model, we use the linear blend skinning (LBS) scheme~\cite{lewis2000pose} 
to compute template mesh deformation:
\begin{equation}
\label{equ:lbs}
\begin{split}
    LBS_{\boldsymbol\theta}(\mathbf{v})=\left(\sum_{j}{\mathbf{w}_{j}(\mathbf{v})\mathbf{T}_{j}(\boldsymbol\theta)}\right)\mathbf{v},
\end{split}
\end{equation}
where $j \leq J$ denotes the index of a joint, $J$ is the number of joints, $\mathbf{T}_{j}(\boldsymbol\theta)$ denotes a $4\times4$ rigid transformation matrix for the $j$-th joint, $\mathbf{w}(\mathbf{v})\in\mathbb{R}^J$ is a skinning weight vector of $\mathbf{v}$ predefined by SMPL model, and $\mathbf{v}$ is homogeneous vertex coordinates of the base mesh. We can conduct articulated deformation by applying \Eq{lbs} to all mesh vertices.
 
The canonical neutral SMPL model $\mathcal{M}_C$ is in the T-pose, and we align the model with each input point cloud $\mathcal{P}_t$ by estimating shape and pose parameters, $\boldsymbol{\beta}$ and $\boldsymbol{\theta}_t$, so that the constructed posed skinned body $\mathcal{M}_{t}$ can fit $\mathcal{P}_t$ well. 
We apply deep virtual markers~\cite{kim2021deep} to $\mathcal{M}_C$ and $\mathcal{P}_t$ to obtain the initial geometric correspondence. We additionally use OpenPose~\cite{cao2019openpose} to improve the point matching accuracy if color images are available.
To obtain the parameters $\boldsymbol{\beta}$ and $\boldsymbol{\theta}_t$ for $\mathcal{M}_t$,
we optimize them using the initial correspondence between $\mathcal{M}_C$ and $\mathcal{P}_t$, and then further minimize the correspondence alignment error and the Chamfer $l_2$ distance together. We add the smoothness regularization term in temporal domain for the optimization so that SMPL's pose parameters can be changed gradually. 

In the face region of the SMPL model, vertices are placed with uneven distribution to provide a detailed face. However, such distribution does not match with the almost uniform point distributions in the input raw scans.
Therefore, we re-mesh the face region of the SMPL model, and the skinning weights of new vertices are assigned from the nearest original vertices.

\subsection{Pose-dependent base mesh}
\label{sec:Method[PBBM]}
Suppose we have obtained a skinned body mesh $\mathcal{M}_{t}$ that approximates the input frame from the previous step. 
To fit the SMPL model tighter to the input point cloud, we combine pose-dependent local deformation with $\mathcal{M}_{t}$ (\Fig{overallProcess}a), and obtain a \textit{pose-dependent base mesh} $\mathcal{B}_{t}$; 
\begin{equation}
\label{equ:pose_dependent_offsets_vert}
    \mathbf{v}' = LBS_{\boldsymbol{\theta}}\left(\mathbf{v} + f_d\left(Q(\mathbf{v}),\overline{\boldsymbol{\theta}}(\mathbf{v})\right)\right),
\end{equation}
where $\mathbf{v}'$ is a vertex of $\mathcal{B}_{t}$.
In \Eq{pose_dependent_offsets_vert}, the displacements are applied to the vertices $\mathbf{v}$ of the T-posed SMPL mesh with the optimized shape parameter $\boldsymbol{\beta}$ before the articulated deformation is performed using LBS function. $f_d$ accounts for pose-dependent local deformation and is implemented as an MLP.
$f_d$ takes as the input a query point $Q(\cdot)$ and a per-point pose feature $\overline{\boldsymbol{\theta}}(\cdot)$ that are described in~\Sec{input}.
We observe such local deformations are useful to handle large shape variations in the input scans.

To optimize $f_d$, we formulate an energy function as follows:
\begin{equation}
\label{equ:pose_dependent_offsets}
\begin{split}
    E_{d}= \sum_t\sum_i\mu_{t,i}\times{d_{C D}\left(\mathbf{p}_{t,i},\mathcal{B}_t\right)} +\lambda_{r} E_{r},
\end{split}
\end{equation}
where $\mathbf{p}_{t,i}\in\mathcal{P}_t$ is a target point in the point cloud at frame $t$, $d_{CD}(A,B)$ evaluates Chamfer distance from $A$ to $B$, and $\mathcal{B}_t$ indicates the base mesh for which SMPL pose parameter $\boldsymbol{\theta}_t$ and pose-dependent local deformation $f_d$ are applied: $\mathcal{B}_t = \mathcal{B}_{\boldsymbol{\theta}_t} = \{\mathcal{V}'_{\boldsymbol{\theta}_t},\mathcal{F}\}$, where $\mathcal{V}'_{\boldsymbol{\theta}_t}=\{\mathbf{v}'_{t,k}\mid k \leq K\}$ and $\mathbf{v}'_{t,k}$ denotes the $k$-th vertex deformed by~\Eq{pose_dependent_offsets_vert} with $\boldsymbol{\theta}_t$.
$\lambda_{r}$ is a weight parameter.
In all equations, for simplicity, we omit averaging terms that divide the sum of Chamfer distances by the number of points.

In~\Eq{pose_dependent_offsets}, per-point weight $\mu_{t,i}$ is used to fully exploit geometric details captured in the input depth images.
Details of nearer objects to the camera are usually better captured than farther ones, and it is advantageous to put more weights on the input points closer to the camera. 
We then use $\mu_{t,i} = e^{-c|z_{t,i}|}$, where $z_{t,i}$ is the depth value of $\mathcal{P}_t$ at frame $t$ and the parameter $c$ is set to 2. 
If the input is a sequence of point clouds, where the distance to the camera is not clear, we use $\mu_{t,i} = 1$.

To avoid noisy artifact, we use Laplacian regularizer $E_{r}$ in \Eq{pose_dependent_offsets} that is defined by
\begin{equation}
\label{equ:base_reg}
    E_{r}=\sum_{t}\sum_{k}{\left|\mathbf{v}'_{t,k}-\frac{\sum_{j\in \mathcal{N}(k)}\mathbf{v}'_{t,j}}{|\mathcal{N}(k)|}\right|^2},
\end{equation}
where $\mathcal{N}(k)$ is the set of adjacent vertices of the $k$-th vertex. $E_{r}$ regularizes the shape of a base mesh $\mathcal{B}_{t}$ to be smooth.
During training time, we construct the total energy by gathering the per-frame energies of randomly sampled frames and optimize the total energy using Adam optimizer~\cite{Adam}. 

Note that, in contrast to DSFN~\cite{burov2021dsfn}, 
we do not conduct any mesh subdivision at this stage for efficiency. 
Indeed, in our experiments, SMPL topology is sufficient to represent a coarse, smooth mesh that is needed for learning neural surface Laplacian function in the following section.

\section{Neural surface Laplacian function}
\label{sec:LapSurfFunc}
To fully exploit the fine-level details in the input point cloud, we construct a neural surface Laplacian function $f_l$ that is an MLP defined on the surface of a pose-dependent base mesh $\mathcal{B}_{t}$.
The input of $f_l$ is the same as for $f_d$, but the output is \emph{approximate Laplacian coordinates}, whereas $f_d$ produces a displacement vector. 

\subsection{Function input}
\label{sec:input}
\begin{figure}[t]
	\begin{tabular}{c c}
		\includegraphics[width=0.194\textwidth]{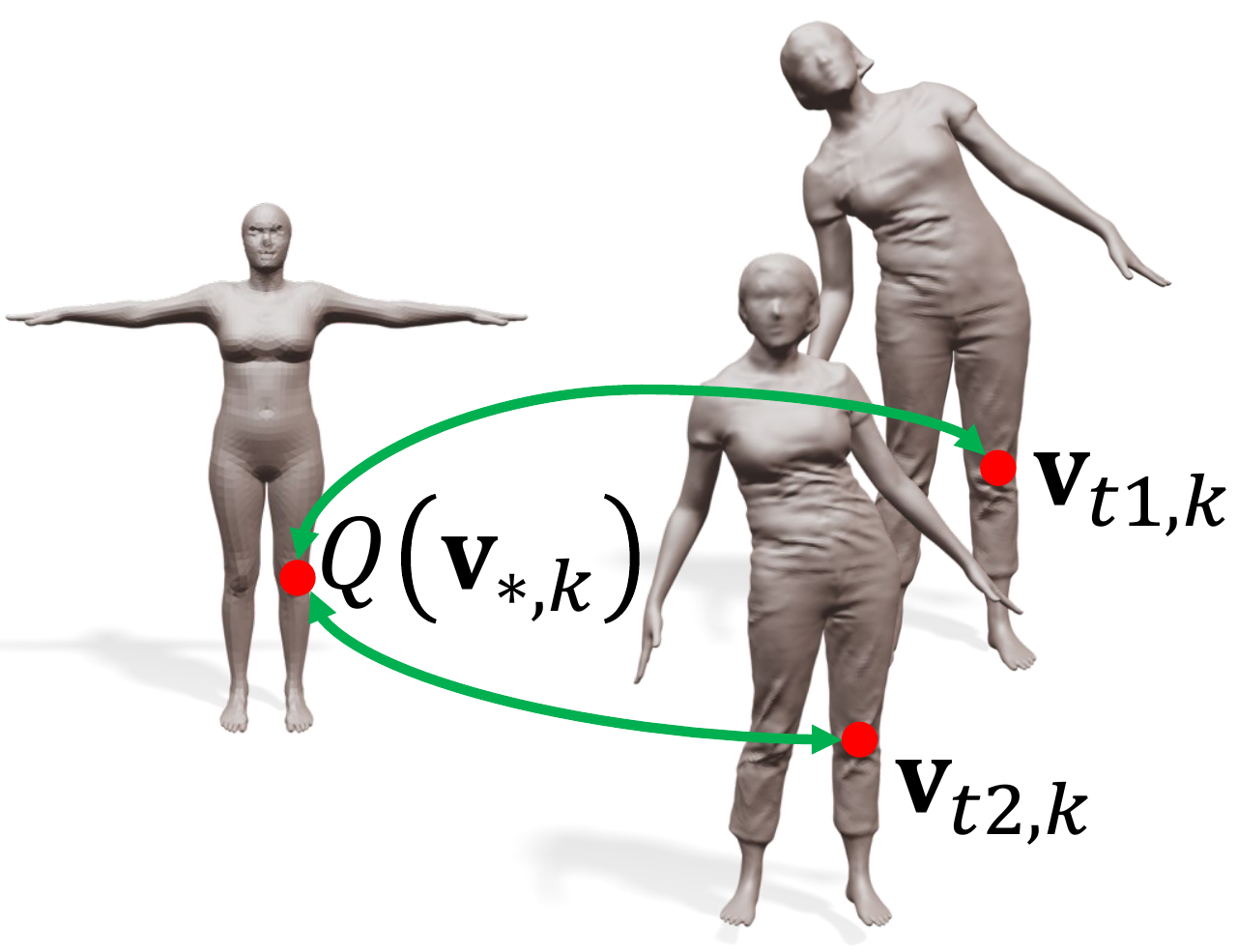} &
		\includegraphics[width=0.26\textwidth]{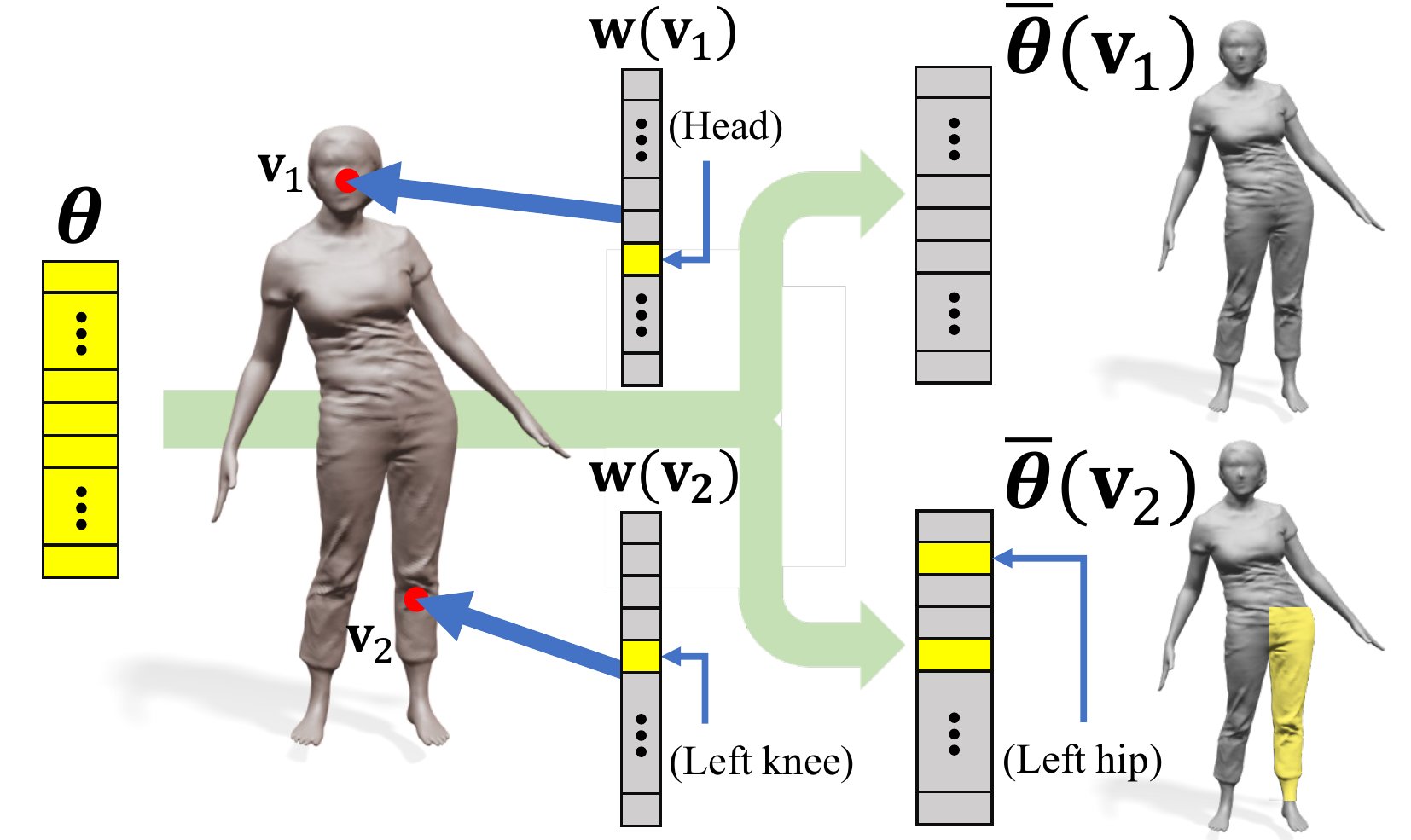} \\
		(a) Query point & (b) Pose feature
	\end{tabular}
	\vspace{-3mm}
	\caption{Input parameters for neural surface Laplacian function. (a) Query points are defined on the canonical neutral SMPL model, so they can be shared among various subjects and different poses. (b) A pose feature is a masked SMPL pose parameter $\boldsymbol{\theta}$ to focus on the relevant joints for a body part. 
	The yellow and gray regions indicate active/inactive parts, respectively.}
	\vspace{-2mm}
	\label{fig:Explanation[input]}
\end{figure}

\paragraph{Query point}
In the inputs of functions $f_d$ and $f_l$,
we use the concept of query point to neutralize shape variations of the base meshes for different subjects and at different frames.
We define a query point $Q(\cdot)$ as a 3D point on the T-posed canonical neutral SMPL model $\mathcal{M}_C$. 
Consider two base meshes $\mathcal{B}_{t1}$ and $\mathcal{B}_{t2}$ for the same subject at different frames and their $k$-th vertices $\mathbf{v}_{t1,k}$ and $\mathbf{v}_{t2,k}$, respectively. 
3D positions of $\mathbf{v}_{t1,k}$ and $\mathbf{v}_{t2,k}$ may differ, but their query points $Q(\mathbf{v}_{t1,k})$ and $Q(\mathbf{v}_{t2,k})$ are defined to be same as they share the same vertex index (\Fig{Explanation[input]}a).
Similarly, $Q(\cdot)$ is defined to be the same for the vertices of base meshes representing different subjects if the vertices have been deformed from the same vertex of $\mathcal{M}_C$.
In addition, $Q(\cdot)$ can be defined for any point on a base mesh other than vertices by using the barycentric coordinates in the mesh triangles.
Once determined, a query point is converted to a high-dimensional vector via a positional encoding $\gamma$~\cite{mildenhall2020nerf}, and we choose the dimension of ten in our experiments.

\paragraph{Pose feature}
For a query point, our two MLPs $f_d$ and $f_l$ should estimate \textit{pose-dependent} deformation and Laplacian coordinates, respectively.
To provide the pose-dependency, we could simply include the pose parameter $\boldsymbol{\theta}$ in the input of the MLPs.
However, a query point is not affected by every joint but strongly associated with nearby joints. For example, the joint angle at the shoulder is irrelevant to local details on a leg.
To exploit the correlations between query points and joint angles in a pose parameter $\boldsymbol{\theta}$, we convert $\boldsymbol{\theta}$ to a per-point pose feature $\overline{\boldsymbol{\theta}}(\mathbf{v})$ that retains only relevant joint angles for a query point $\mathbf{v}$. 
Inspired by Pose Map~\cite{saito2021scanimate}, we apply the joint association weight map $\mathbf{W}\in\mathbb{R}^{J\times J}$ and skinning weights $\mathbf{w}(\mathbf{v})\in\mathbb{R}^J$ of $\mathbf{v}$ to the original pose parameter $\boldsymbol{\theta}\in\mathbb{R}^{J\times3}$. Our pose feature used as the input of MLPs is defined by
\begin{equation}
\label{equ:pose_feature}
    \overline{\boldsymbol{\theta}}(\mathbf{v}) = \lceil diag \left(\mathbf{W}\,\mathbf{w}(\mathbf{v}) \right) \rceil {\boldsymbol{\theta}},
\end{equation}
where $diag(\cdot)$ converts an input vector to a diagonal matrix, and $\lceil\cdot\rceil$ is an element-wise ceiling operation.
We manually define the weight map $\mathbf{W}$ to reflect our setting. For example, the details of the head are not correlated with any joint in our reconstructed model, and the details of a leg are affected by all nearby joints together. Then, for the head joint, we set zero for the association weight of any joint in $\mathbf{W}$. For a joint around a leg, we set higher association weights for all nearby joints (\Fig{Explanation[input]}b).

\subsection{Training pairs}
\label{sec:train_pairs}
To train the neural surface Laplacian function $f_l$, we calculate the ground-truth (GT) approximate Laplacian coordinates of scan points and localize them on $\mathcal{M}_C$ to the corresponding query points.

\begin{figure}[t]
	\begin{tabular}{c c c}
		\includegraphics[width=0.07\textwidth]{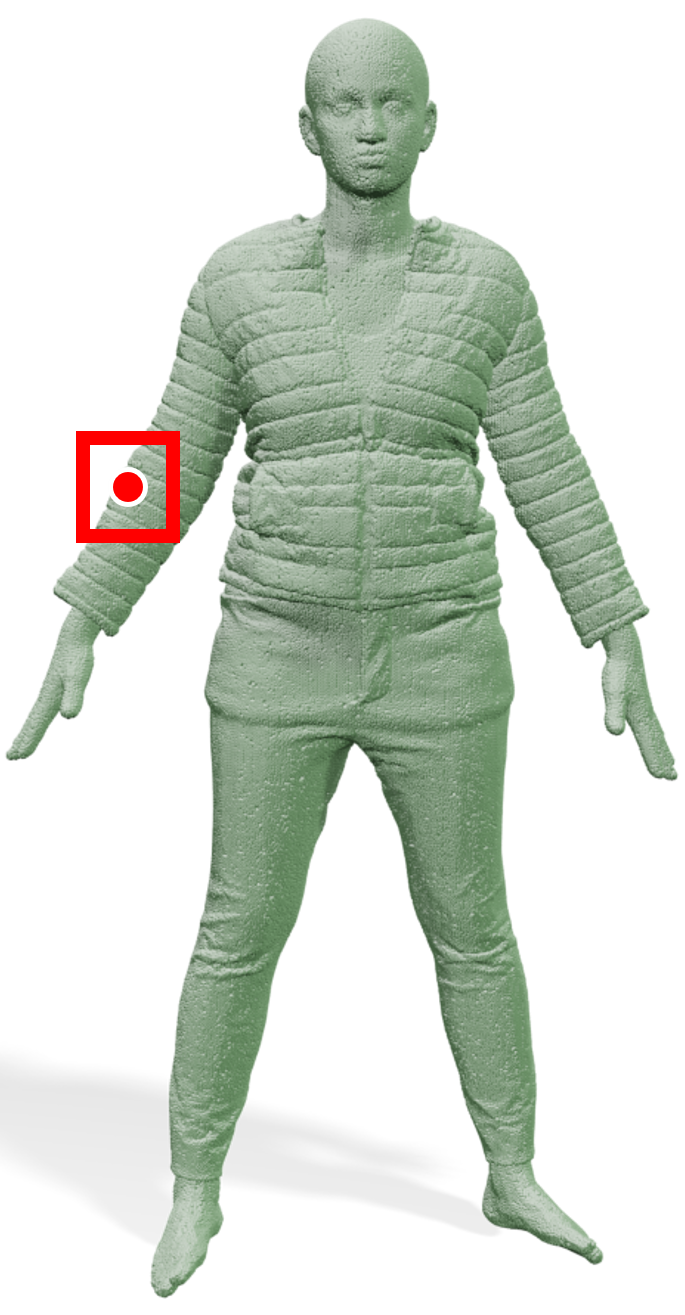} &
		\includegraphics[width=0.15\textwidth]{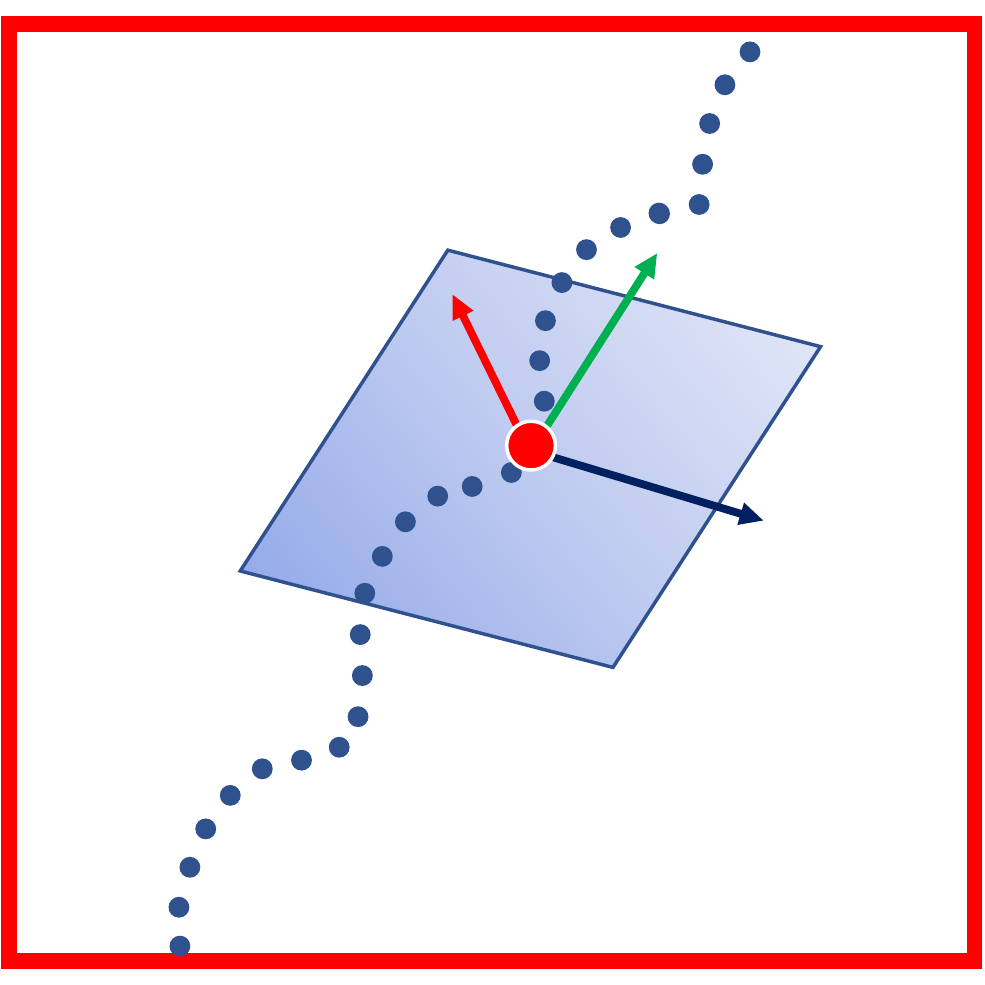} &
		\includegraphics[width=0.15\textwidth]{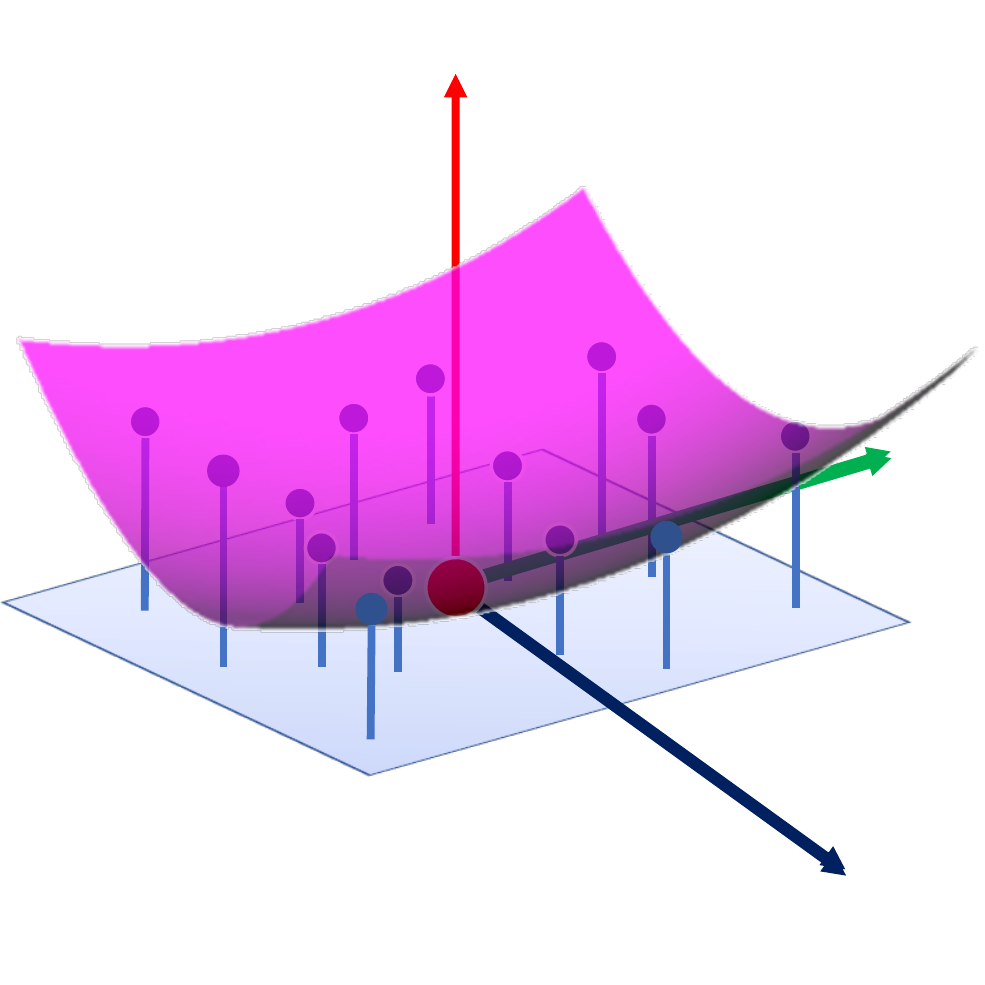} \\
		(a) Scan & (b) Local coordinate system & (c) Quadric surface \\
	\end{tabular}
	\vspace{-2mm}
	\caption{Illustration of GT Laplacian coordinates approximation. To compute Laplacian coordinates on a point cloud (a), we initially define a local coordinate system (b), and compute a quadratic polynomial that locally fits the point cloud. Then, the coefficients of the quadratic polynomial are used to obtain Laplacian coordinates. }
	\vspace{-3mm}
	\label{fig:Explanation[lap_approx]}
\end{figure}

\paragraph{GT Laplacian coordinates approximation}
As discussed in Section \ref{sec:LapPCD}, we use an approximation method~\cite{liang2012geometric} for calculating Laplacian coordinates from scan points that do not have connectivity. We first locally fit a degree two polynomial surface for each point in the moving least squares manner. 
In our experiments, we use 20-30 neighbor points for the local surface fitting. 
\Fig{Explanation[lap_approx]} shows illustration.
Our approximate Laplacian coordinates are defined on a continuous domain, unlike in conventional mesh editing methods~\cite{lipman2004differential, sorkine2004laplacian} that formulate Laplacian coordinates in a discrete domain using mesh vertices. To apply our Laplacian coordinates to a discrete mesh, we take advantage of a pose-dependent base mesh
(\Sec{Method[Recon]}).

\paragraph{Localization}
Although the base mesh $\mathcal{B}_t$ nearly fits the input point cloud, points may not reside exactly on the mesh. To obtain the corresponding query point, we project each point $\mathbf{p}$ in the input point cloud $\mathbf{P}_t$ onto the base mesh $\mathcal{B}_t$:
$\overline{\mathbf{p}}=\Pi\left(\mathcal{B}_t,\mathbf{p}\right)$,
where $\Pi$ denotes the projection operation from a point to a pose-dependent base mesh
(\Fig{overallProcess}b).
The position of $\,\overline{\mathbf{p}}_{t,i}$ is determined by the barycentric coordinates in a triangle of $\mathcal{B}_t$, so we can easily compute the query point, skinning weights, and pose feature using the barycentric weights.

\subsection{Optimization}
\label{sec:optimization}
\paragraph{neural surface Laplacian function}
For a given surface point $\overline{\mathbf{p}}$ and pose $\boldsymbol{\theta}$, the MLP $f_l$ estimates Laplacian coordinates:
\begin{equation}
    \boldsymbol{\delta}'(\overline{\mathbf{p}})=LBS_{\boldsymbol{\theta}}\left(f_l\left(Q(\overline{\mathbf{p}}), \overline{\boldsymbol{\theta}}(\overline{\mathbf{p}})\right)\right).
\label{equ:pose_dependent_laplacian}    
\end{equation}
The estimation is conducted in the canonical space and transformed into the posed space. 
Working in the canonical space is essential because Laplacian coordinates are not invariant to rotation. 
In~\Eq{pose_dependent_laplacian}, we discard the translation part in~\Eq{lbs} as Laplacian coordinates are differential quantity.
The estimated $\boldsymbol{\delta}'$ is non-uniform Laplacian coordinates, as described in~\Sec{LapPCD}.

We train the MLP $f_l$ by formulating a per-point energy function:
\begin{equation}
    E_{l}=\sum_{t}\sum_{i}\mu_{t,i}{\left|\boldsymbol{\delta}'_{t,i}-\boldsymbol{\delta}_{t,i} \right|^2},
\label{equ:PDD}
\end{equation}
where $\boldsymbol{\delta}'_{t,i}$ is the Laplacian coordinates of $\overline{\mathbf{p}}_{t,i}$ predicted by $f_l$ using~\Eq{pose_dependent_laplacian}, $\boldsymbol{\delta}_{t,i}$ is the GT approximate Laplacian coordinates of $\mathbf{p}_{t,i}$, and $\mu_{t,i}$ is the weight used in~\Eq{pose_dependent_offsets}.
During training, we construct the total energy by summing per-point energies of randomly sampled input points, and optimize the total energy using Adam optimizer~\cite{Adam}.

\section{Laplacian Reconstruction}
\label{sec:Method[Recon]}
Once the training is over, for a given pose $\boldsymbol{\theta}$, we can obtain the pose-dependent base mesh $\mathcal{B}_{\boldsymbol{\theta}}$ and the Laplacian coordinates for the vertices of $\mathcal{B}_{\boldsymbol{\theta}}$. In the reconstruction step, we aggregate the estimated Laplacian coordinates to restore a whole body model.

\paragraph{Subdivision}
For detailed surface reconstruction, 
we subdivide the base mesh $\mathcal{B}_{\boldsymbol{\theta}}$ as the SMPL model does not have enough number of vertices for representing fine details. 
The new vertices of the subdivided mesh $\mathcal{B}$ reside on the midpoints of edges of $\mathcal{B}_{\boldsymbol{\theta}}$. We conduct subdivision twice, and the number of triangles increases 16 times. 
As a result, we have the subdivided pose-dependent base mesh $\mathcal{B}=\{\mathcal{U},\mathcal{F}'\}$,
where $\mathcal{U}=\{\mathbf{u}_{k}\mid k \leq K'\}$ and $K'$ is the number of vertices of $\mathcal{B}$.

\paragraph{Reconstruction}
\label{sec:reconstruction}
Using the base mesh $\mathcal{B}$, as illustrated in \Fig{overallProcess}c, we reconstruct a detailed mesh $\mathcal{S}=\left(\mathcal{U}',\mathcal{F}'\right)$, where $\mathcal{U}'=\{\mathbf{u}'_k\mid k\leq K'\}$, by minimizing the following error functional~\cite{lipman2004differential, sorkine2004laplacian}:
\begin{equation}
    E(\mathcal{U}')=\sum_{k}{\left\| \Delta_{\mathcal{B}}\left(\mathbf{u}'_k\right)-\boldsymbol{\delta}'\left(\mathbf{u}_k\right)\right\|^2}+\sum_{k\in \:anchor}{\left\| \mathbf{u}'_k - \mathbf{u}_k \right\|^2},
\label{equ:LapRecon}
\end{equation}
where 
$\Delta_{\mathcal{B}}$ is the Laplace-Beltrami operator of $\mathcal{B}$ defined in~\Eq{cotLap}, 
$\boldsymbol{\delta}'(\mathbf{u}_k)$ is the non-uniform Laplacian coordinates of $\mathbf{u}_k$ predicted by the neural surface Laplacian function $f_l$ 
using~\Eq{pose_dependent_laplacian}, 
and $anchor$ is a set of indices of the constraint vertices on $\mathcal{B}$, which play the role of boundary conditions.
In \Eq{LapRecon}, the first term preserves desirable Laplacian coordinates on the whole surface, and the second term constrains the global position of the final shape using the anchor points. 

\paragraph{Laplace-Beltrami operator}
To compute the Laplace-Beltrami operator $\Delta_{\mathcal{B}}$ using~\Eq{cotLap}, 
we need the angles $\alpha^1_{k,j}$ and $\alpha^2_{k,j}$ for each edge connecting vertices $\mathbf{u}_k$ and $\mathbf{u}_j$ in $\mathcal{B}$.
For efficient computation, we use uniform angle $\alpha=\alpha^1=\alpha^2=\frac{\pi}{2}-\frac{\pi}{|\mathcal{N}(k)|}$ for all edges of $\mathcal{B}$.
Then, \Eq{cotLap} is reduced to 
\begin{equation}
\label{equ:approxLB}
\begin{split}
    \Delta_{\mathcal{B}} \left(\mathbf{u}'_k\right)&=\frac{cot(\alpha)}{a_k}
    \left(|\mathcal{N}(k)|\;\mathbf{u}'_k-\sum_{j\in \mathcal{N}(k)}{\mathbf{u}'_j}\right),
\end{split}
\end{equation}
where $a_k$ is the Voronoi area of a vertex $\mathbf{u}_k$ of $\mathcal{B}$. 
As shown in the experimental results (\Sec{exp}), this approximation successfully reconstructs $\mathcal{S}$ from the predicted non-uniform Laplacian coordinates.

\begin{figure}[t]
	\centering
	\includegraphics[width=0.47\textwidth]{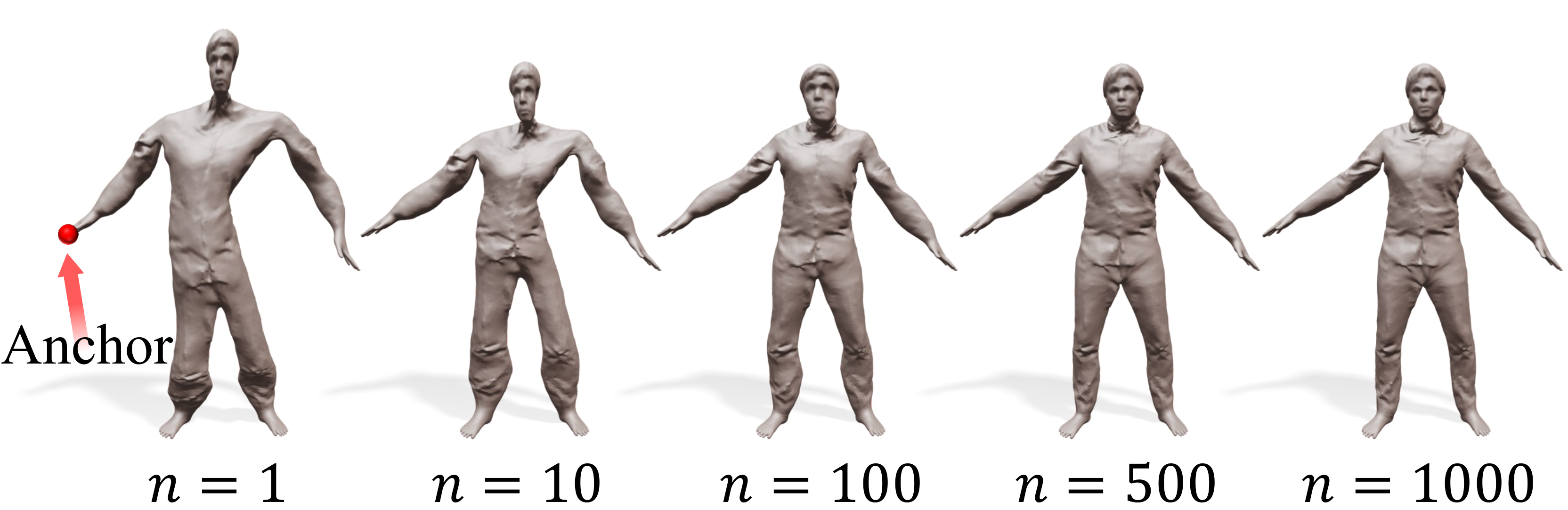}
	\vspace{-2mm}
	\caption{Effect of the number of anchor points. On the left, the red dot represents a single anchor point used for reconstruction. Too few anchor points ($n\leq500$) introduce distortions in the reconstruction results. 
	}
	\label{fig:ablation[anchor_points]}
\end{figure}

\paragraph{Anchor points}
Theoretically, we are able to recover the original mesh from the Laplacian coordinates of vertices by fixing one vertex position as an anchor and solving a linear system~\cite{sorkine2004laplacian}.
However, in our setting, one anchor may not suffice for reconstructing the accurate final surface $\mathcal{S}$, as the Laplacian coordinates are approximate ones predicted by a neural surface Laplacian function $f_l$, not computed directly from $\mathcal{S}$.
To improve the accuracy of reconstruction, we set anchor points of a sufficient number as the boundary condition.
We select a set of $n$ vertex indices as the $anchor$ in \Eq{LapRecon} in advance by uniformly sampling~\cite{yuksel2015sample} vertices from the canonical neutral SMPL model $\mathcal{M}_C$, where $n=800$ in our experiments. 
\Fig{ablation[anchor_points]} shows reconstruction results with varying numbers of anchor points.

\paragraph{Reliable anchors}
Since anchor points are fixed for solving~\Eq{LapRecon}, they should be as close to the input point cloud as possible. To achieve this property, we set an additional energy term and optimize it along with \Eq{pose_dependent_offsets} when we train the pose-dependent local deformation function $f_d$:
\begin{equation}
\label{equ:reliable_anchor}
\begin{split}
    E_{a} = \lambda_{a}\sum_t\sum_{k \in \;anchor}d_{CD}\left(\mathbf{v}'_{t,k},\mathcal{P}_t\right),
\end{split}
\end{equation}
where $d_{CD}$ measures a vertex-to-point cloud Chamfer distance, $\mathcal{P}_t$ is the input point cloud at frame $t$, and $\lambda_{a}$ is a weight parameter.
When the input is a depth map sequence, we apply this term to only visible anchors from the camera viewpoint.

\section{Results}
\label{sec:exp}
\subsection{Experiment details}

\begin{figure}[t]
	\centering
	\includegraphics[width=0.45\textwidth]{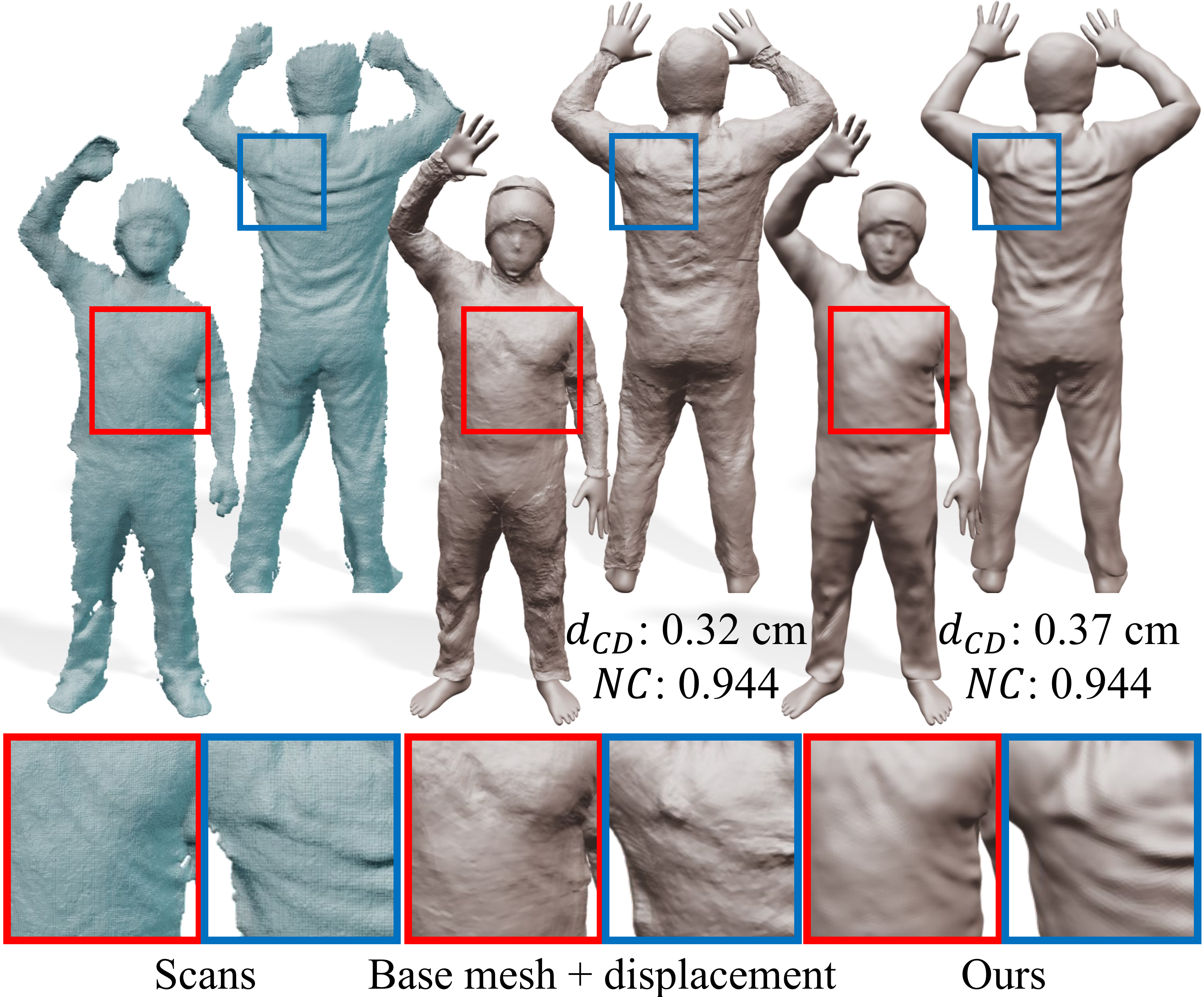}\\
	\vspace{-2mm}
	\caption{Comparison with using a regularization-free displacement function on the base mesh. Although the surface displacement function is trained in the same condition as our neural surface Laplacian function, it cannot capture structure details (middle). In contrast, our results (right) successfully restore details from the input scan (left). 
	}
	\label{fig:ablation[pose_dependent_details]}
\end{figure}

\paragraph{Implementation and training details}
Pose-dependent local deformation function $f_d$ and neural surface Laplacian function $f_l$ are represented as 5- and 3-layer MLPs with ReLU activation, and 600 and 800 feature channels are used per intermediate layer, respectively.
In our experiments, we set $\lambda_r=0.1\sim 2$ and $\lambda_a=2$.
We optimize $f_d$ and $f_l$ with a learning rate of $1.0\times 10^{-3}$, batch sizes of $10$ frames and 5000 points, and 300 and 100 epochs, respectively.
The training time is proportional to the numbers of points and frames in the scan data. For instance, the point cloud sequence used for reconstruction in~\Fig{ablation[pose_dependent_details]} consists of 39k$\sim$56k points per frame with 200 frames. In that case, it takes about 20 and 30 minutes to train MLPs $f_d$ and $f_l$, respectively.

\paragraph{Datasets}
We evaluate the results of our method qualitatively and quantitatively on single-view and full-body point cloud sequences.
We capture RGB-D sequences using an Azure Kinect DK~\cite{KinectAzure} with 1280p resolution for color images and $1024 \times 1024$ for depth images. Additionally, we use RGB-D sequences that are provided in DSFN~\cite{burov2021dsfn}. RGB-D sequences used for experiments contain 200 to 500 frames. We also evaluate our method on full-body point clouds using synthetic datasets: CAFE~\cite{ma2020cape}, and Resynth~\cite{POP:ICCV:2021}.
We show the reconstruction result from a 2.5D depth point cloud in \textit{mint color} and the result of a full-body 3D point cloud in \textit{olive color}.

\paragraph{Timings}
In the reconstruction step, for the example in~\Fig{ablation[pose_dependent_details]}, it takes 3ms and 35ms per frame to evaluate the MLPs $f_d$ and $f_l$, respectively.
For solving a sparse linear system to obtain the final detailed mesh $\mathcal{S}$ that minimizes \Eq{LapRecon},
it takes 4 seconds for the matrix pre-factorization step and 130ms to obtain the solution for each frame when the mesh contains 113k vertices.

\subsection{Analysis}
\paragraph{Effect of Laplacian coordinates}

Our approach using Laplacian coordinates preserves local geometric details better than the approach using absolute coordinates of mesh vertices. To verify the claim, we conduct an experiment by changing the neural surface Laplacian function to estimate displacements instead of Laplacian coordinates. We think that this surface function mimics a pose-dependent displacement map.
To optimize the surface displacement function, we use the same energy function as \Eq{PDD} with the change of $\boldsymbol{\delta}$ to the displacement between surface point $\overline{\mathbf{p}}$ and scan point $\mathbf{p}$.
There is no regularization term in \Eq{PDD}, and the maximal capability of the displacement function is exploited for encoding surface details.
Nevertheless, the resulting displacement function cannot properly capture structure details, producing rather noisy surfaces (\Fig{ablation[pose_dependent_details]} middle).  
In contrast, our results using Laplacian coordinates are capable of capturing structure details (\Fig{ablation[pose_dependent_details]} right) from the input point clouds (\Fig{ablation[pose_dependent_details]} left). 
In~\Fig{ablation[pose_dependent_details]}, we include quantitative results (Chamfer distance $d_{CD}$ and normal consistancy $NC$) on our dataset. The normal consistency $NC$ is computed using the inner products of the ground truth normals at input points and the normals of the corresponding points on the reconstructed surface. Our result shows a slightly higher Chamfer distance error than the displacement function, but produces more visually pleasing results. 

\begin{figure}[t]
	\centering
    \includegraphics[width=0.47\textwidth]{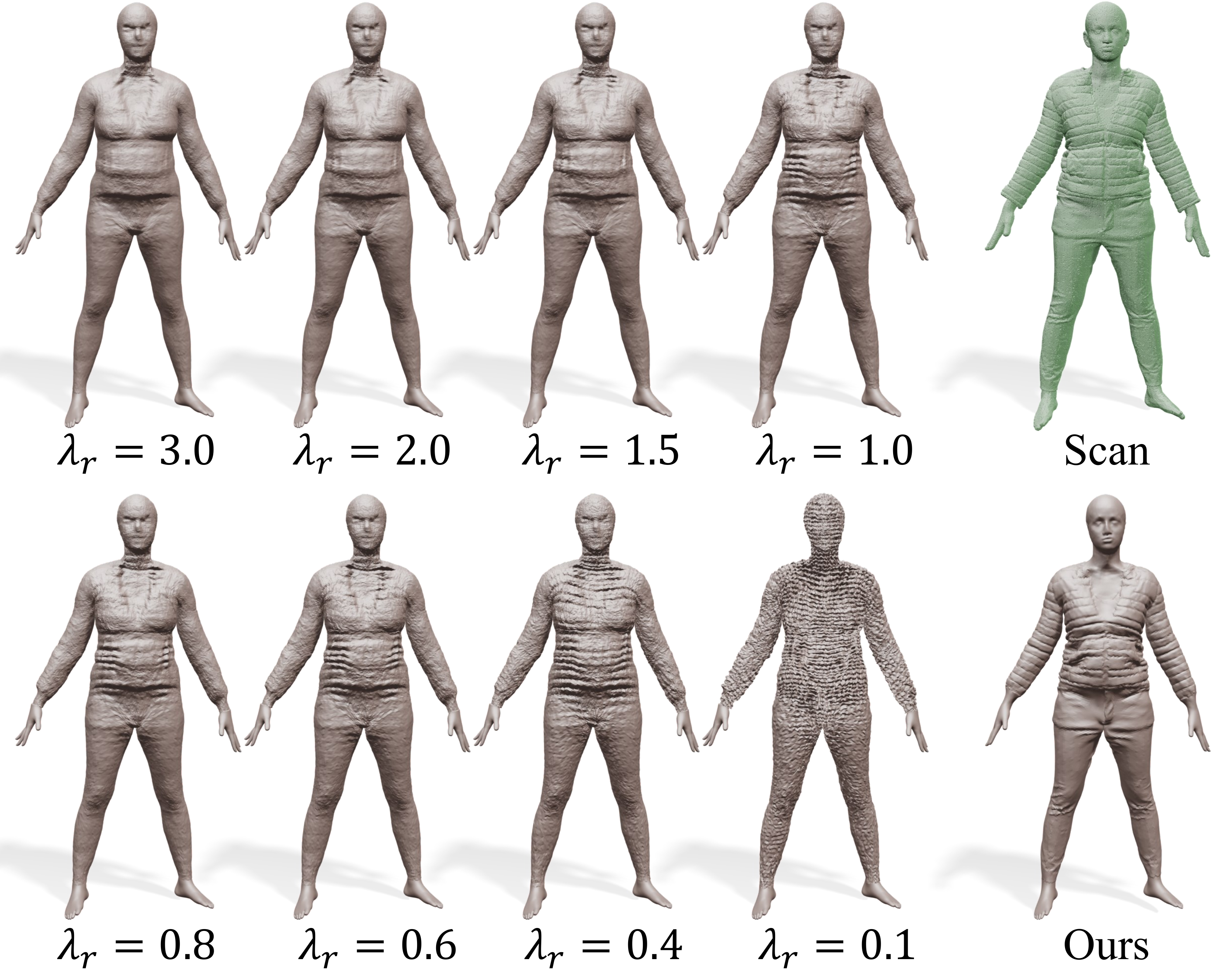} \\
	\vspace{-3mm}
    \caption{Pose-dependent local deformation (Sec.~\ref{sec:Method[PBBM]}) applied to the subdivided base mesh used for our final reconstruction. (left) We train the deformation function with varying regularization weights. It is hard to find the best weight for regularization as the structures and sizes of shape details are spatially varying in the raw scan (top right). (bottom right) The reconstruction result from our complete pipeline restores shape details with appropriate structures and sizes.
	}
	\vspace{-6mm}
	\label{fig:ablation[onlyposeblend]}
\end{figure}

\begin{figure}[t]
	\centering
	\includegraphics[width=0.47\textwidth]{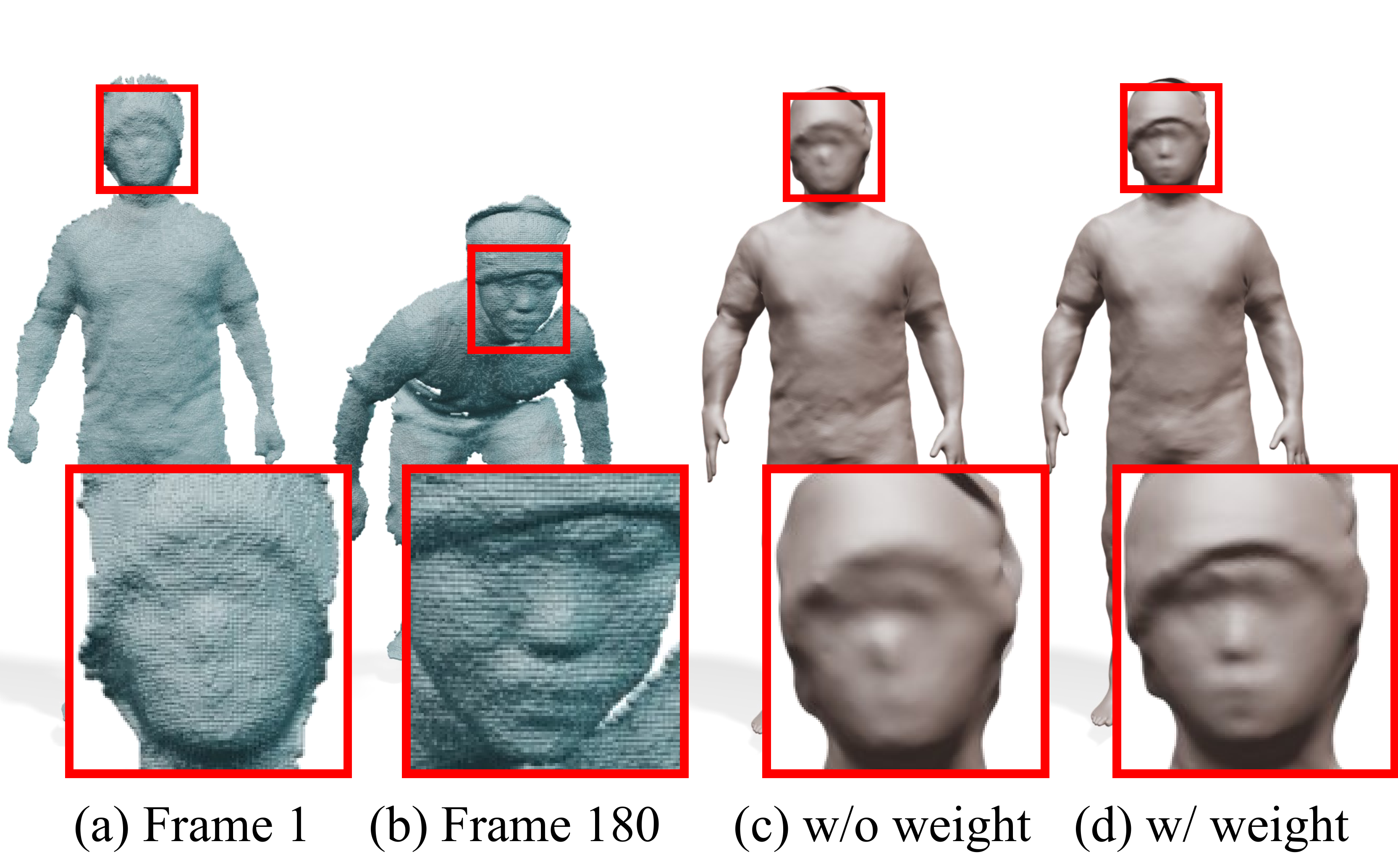} \\
	\vspace{-3mm}
	\caption{Effect of selective weighting. Without selective weighting (c), surface functions may learn geometric information from low-quality frames (a). Our selective weighting (d) encourages surface functions to learn sharp geometric details (b) in the input depth map sequence.
	}
	\label{fig:ablation[selective_weighting]}
\end{figure}

\begin{figure}[t]
	\centering
	\begin{tabular}{c}
	   \includegraphics[width=0.45\textwidth]{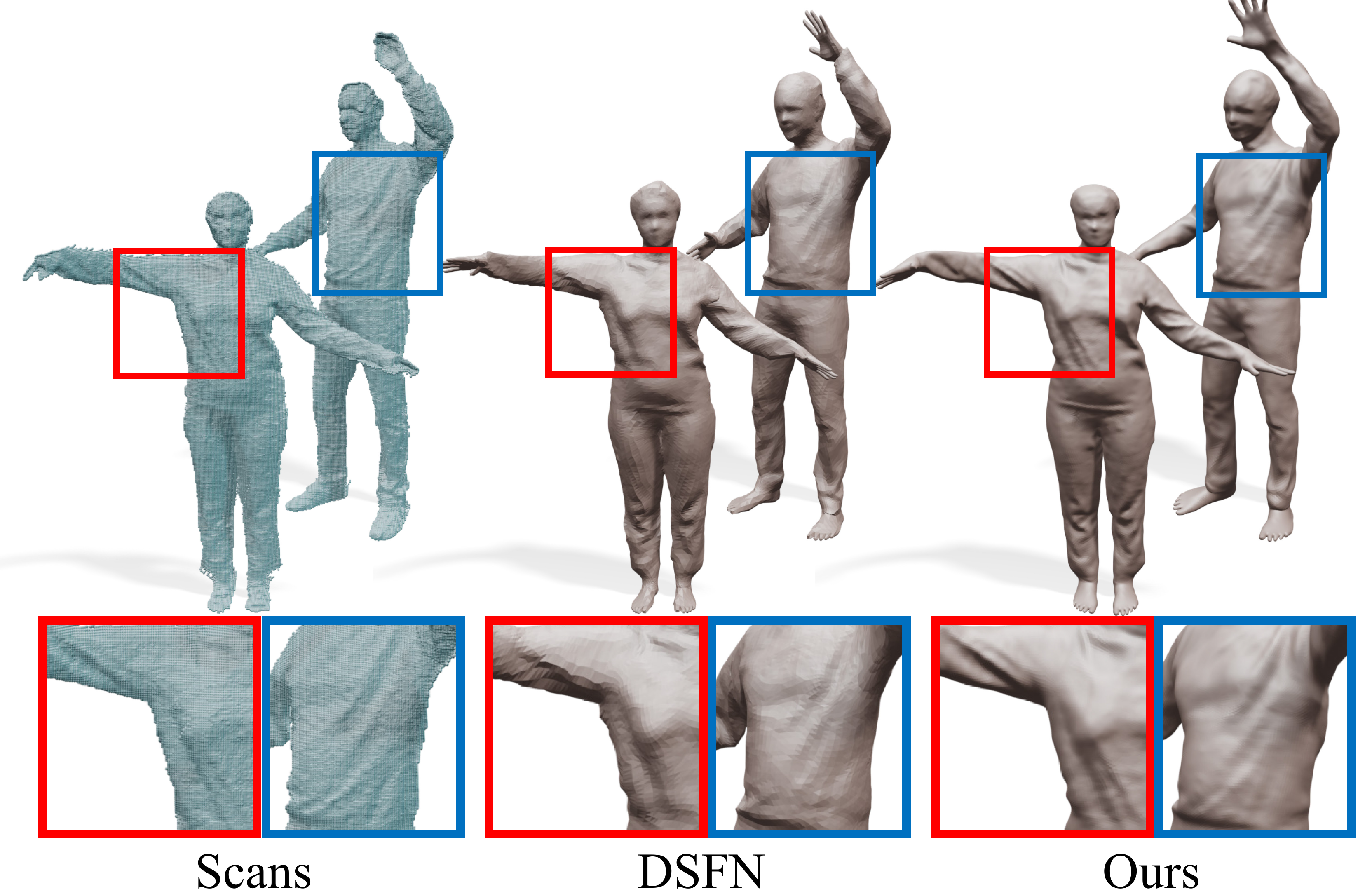} \\
	\end{tabular}
	\vspace{-2mm}
	\caption{Comparison with DSFN~\cite{burov2021dsfn} on DSFN real dataset captured by a single RGB-D camera.
	}
	\vspace{-3mm}
	\label{fig:Comparison[DSFN1]}
\end{figure}

\paragraph{Unclear best weight for smoothness regularization on local deformation}
We conduct an experiment that learns the local deformation defined in Sec.~\ref{sec:Method[PBBM]} for a subdivided base mesh that has the same topology as our final mesh. The results (\Fig{ablation[onlyposeblend]} left) show that estimating absolute coordinates is sensitive to the regularization weight $\lambda_r$ defined in~\Eq{pose_dependent_offsets}. With a large value of $\lambda_r$, shape details are smoothed out in the reconstructed model. With a small $\lambda_r$, too small details are reconstructed. Then, it is hard to find the best regularization weight to handle spatially varying structures and sizes of shape details in the input scan (\Fig{ablation[onlyposeblend]} top right). On the contrary, our complete pipeline that learns differential representation performs better without a regularization term (\Fig{ablation[onlyposeblend]} bottom right). Note that all meshes in \Fig{ablation[onlyposeblend]} has the same number of vertices.

\paragraph{Effect of selective weighting}
We use selective weighting $\mu_{t,i}$ in \Eq{pose_dependent_offsets} and \Eq{PDD} to obtain the best details from the input depth map sequence. This weighting is especially effective in the face region (\Fig{ablation[selective_weighting]}).

\subsection{Comparisons}
We compare our LaplacianFusion with recent representations developed for \textit{controllable} 3D clothed human reconstruction. The compared representations are based on mesh~\cite{burov2021dsfn}, signed distance function (SDF)~\cite{wang2021metaavatar}, and point cloud~\cite{POP:ICCV:2021}.

\begin{figure}[t]
	   \includegraphics[width=0.45\textwidth]{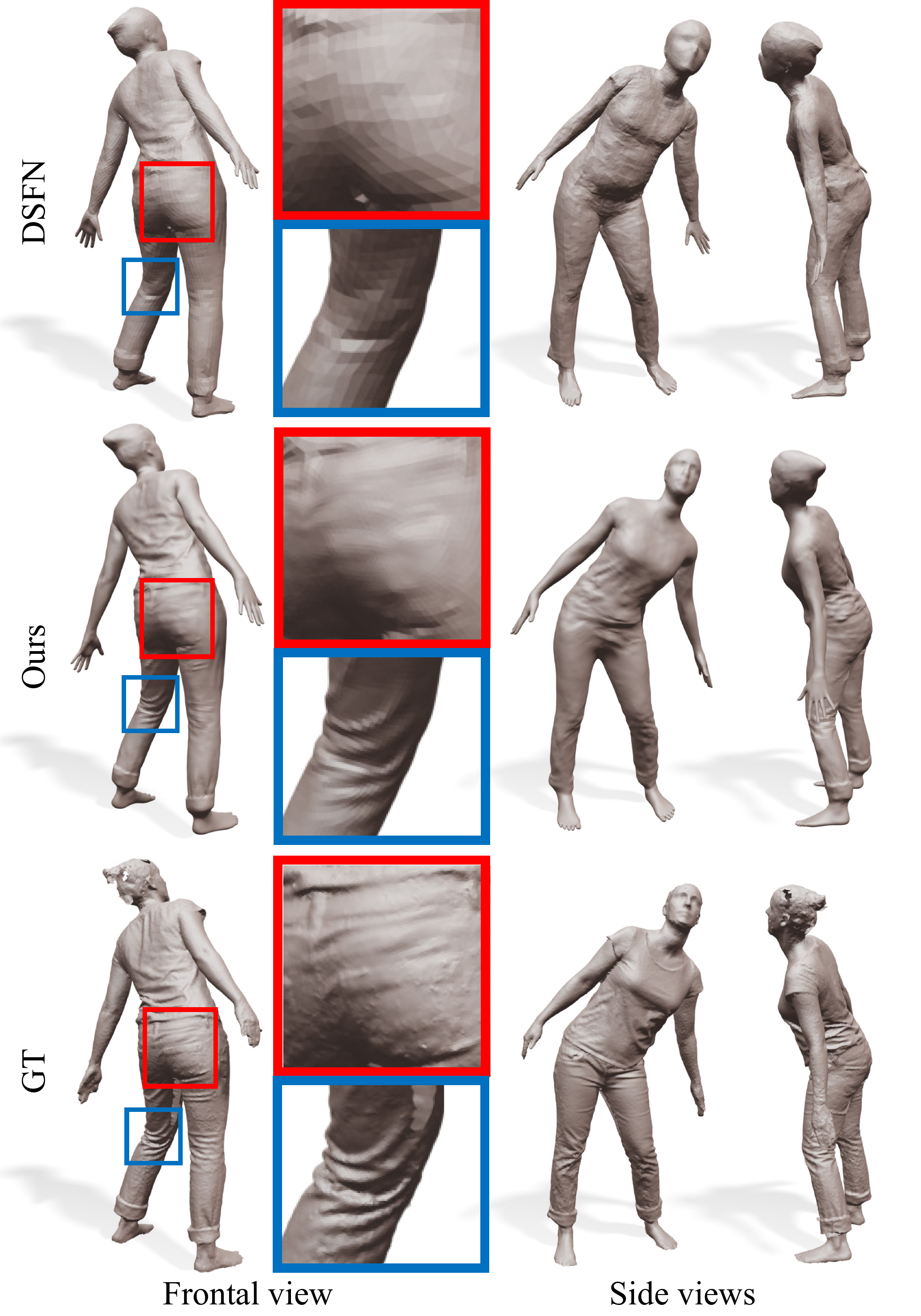} \\
	\vspace{-3mm}
	\caption{Comparison with DSFN~\cite{burov2021dsfn} on BUFF dataset~\cite{Zhang_2017_CVPR}. Note that the training input is not a ground truth sequence (bottom row) but a rendered RGB-D sequence with $640\times 480$ resolution. The red and blue boxes highlight qualitative differences.
	}
	\label{fig:Comparison[DSFN]}
\end{figure}

\paragraph{Comparison with DSFN (Mesh)}
DSFN~\cite{burov2021dsfn} utilizes an explicit mesh to reconstruct a controllable 3D clothed human model from a RGB-D sequence. It represents local details with an offset surface function (dense local deformation), but its spatial regularization smooths out geometric details. Since the source code of DSFN is not published, we compared our results with DSFN on the dataset provided by the authors.

Their real data inputs are captured using an Azure Kinect DK~\cite{KinectAzure} with $1920 \times 1080$ pix.\ for color images, and $640 \times 576$ pix.\ for depth images. \Fig{Comparison[DSFN1]} shows results on the real dataset provided by DSFN, and our approach produces more plausible details than DSFN. Note that, in this dataset, the captured human stands away from the camera, so input point cloud lacks high-frequency details.

For quantitative evaluation, DSFN uses synthetic $640 \times 480$ pix.\ RGB-D sequences rendered from BUFF~\cite{Zhang_2017_CVPR}, where the camera moves around a subject to cover the whole body.
\Fig{Comparison[DSFN]} shows the comparison results, and our method preserves the details of the input better.
\Tbl{DSFN} shows the measured IOU, chamfer distance ($d_{CD}$), normal consistency ($NC$) scores. All three scores of our approach rate better than DSFN.

\begin{table}[t]
    \centering
    \caption{Quantitative comparisons with DSFN~\cite{burov2021dsfn} using a sequence of the BUFF dataset~\cite{Zhang_2017_CVPR}.}
    \resizebox{.47\textwidth}{!}{
    \begin{tabular}{c| c| c| c}
        \hline
        Method & IoU $\uparrow$ & $d_{CD}$(cm) $\downarrow$ & $NC$ $\uparrow$\\
         \hline
        DSFN~\shortcite{burov2021dsfn} & 0.832 & 1.56 & 0.917 \\
         \hline
         Ours (same vertex density as DSFN) & 0.863 & 0.99 & 0.933 \\
         \hline
         Ours & \textbf{0.871} & \textbf{0.94} & \textbf{0.941} \\
         \hline
    \end{tabular}
    }
	\vspace{-2mm}
    \label{tbl:DSFN}
\end{table}

\begin{figure}[t]
	    \includegraphics[width=0.45\textwidth]{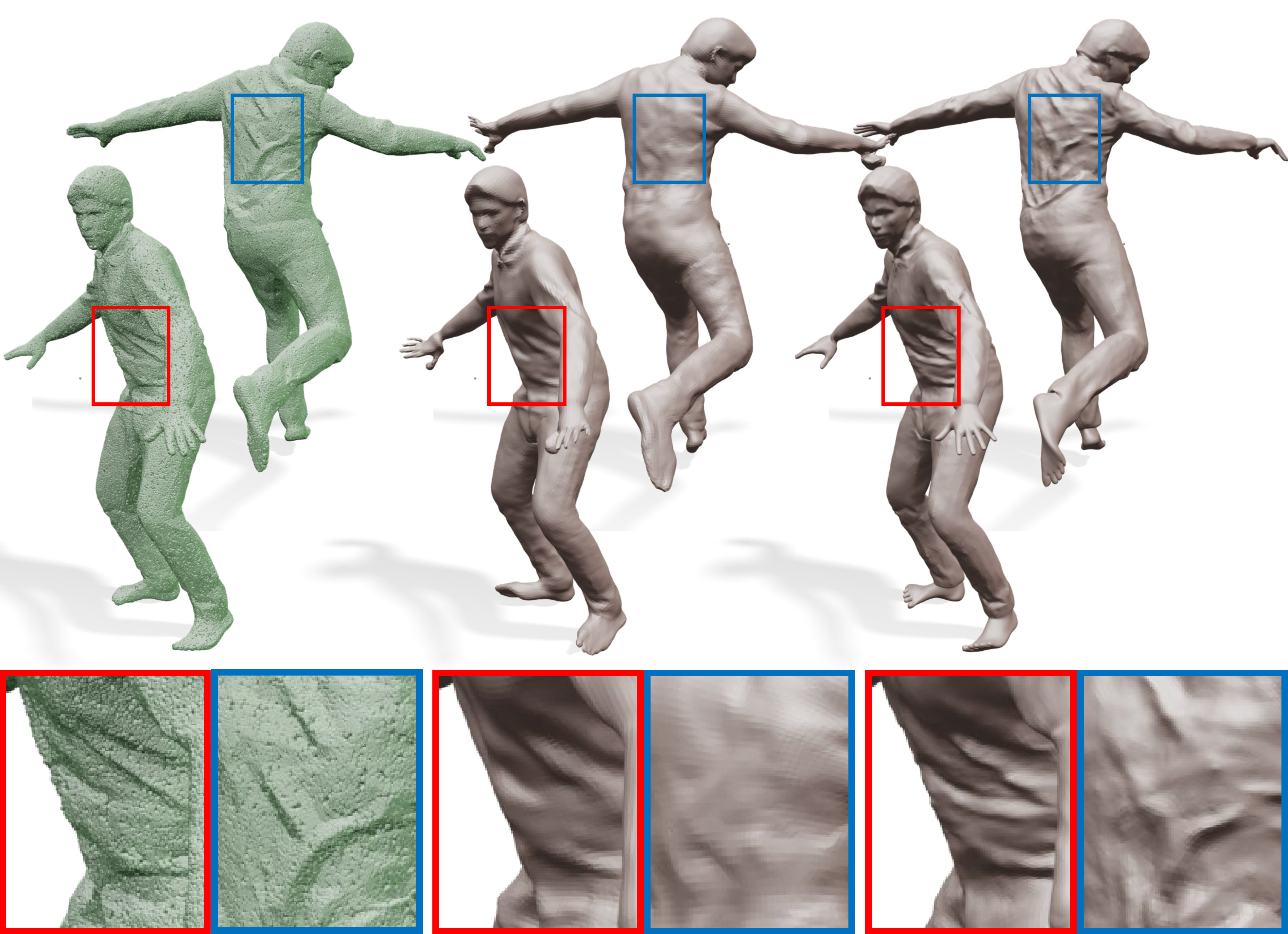}\\
	    \includegraphics[width=0.45\textwidth]{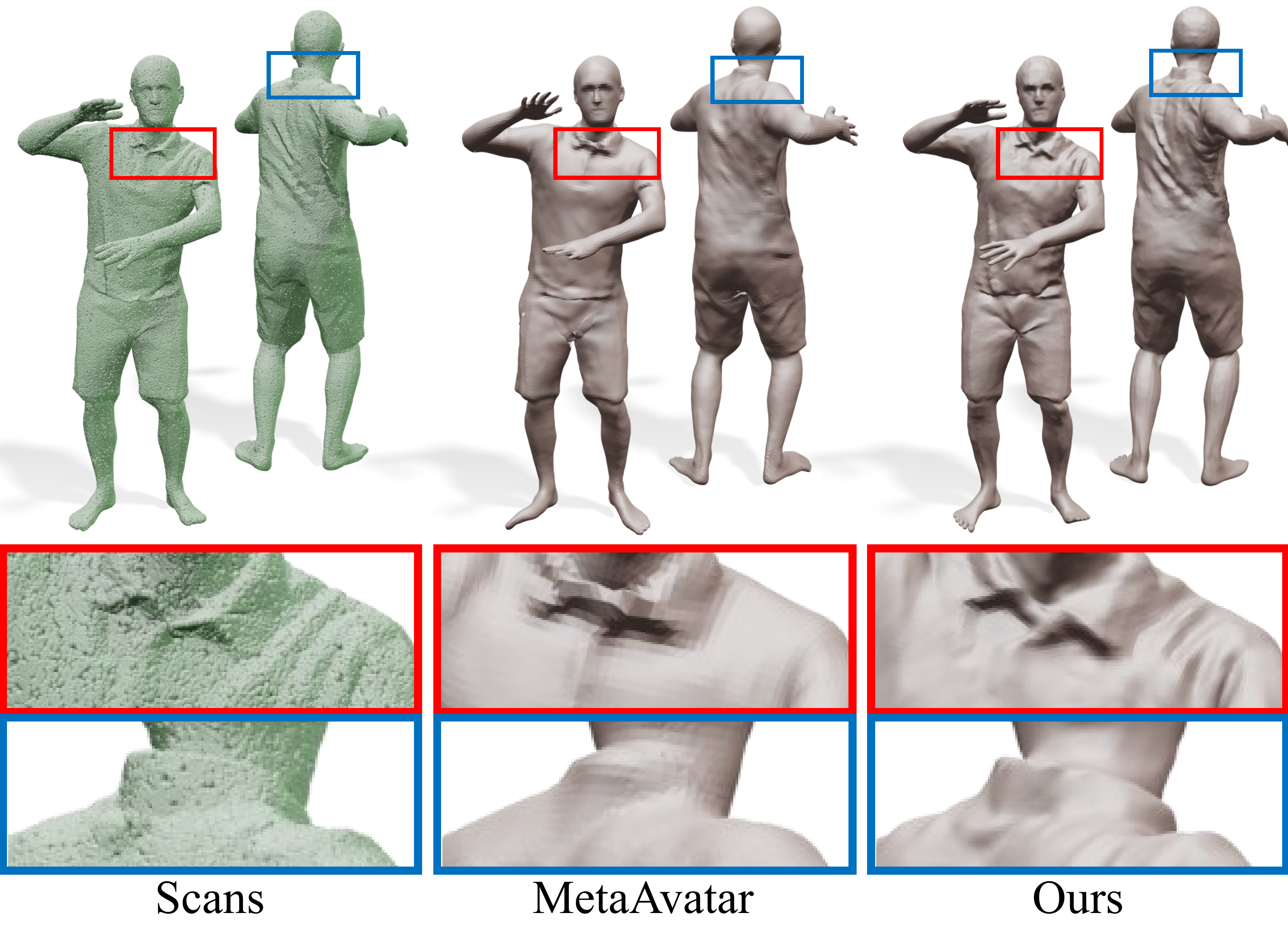}\\
	\vspace{-3mm}
	\caption{Comparison with MetaAvatar~\cite{wang2021metaavatar} on CAPE dataset~\cite{ma2020cape}. (From left to right) raw 3D point clouds, results of MetaAvatar, and our results.
	}
	\vspace{-3mm}
	\label{fig:Comparison[Meta]}
\end{figure}

\paragraph{Comparison with MetaAvatar (SDF)}
Wang et al.~\shortcite{wang2021metaavatar} propose MetaAvatar that represents shape as SDF defined on 3D space. The resulting shape tends to be smoothed due to the limited number of sampling points in the training time. MetaAvatar utilizes the given SMPL parameters for reconstruction, and we use them in the comparison.
MetaAvatar evaluates the reconstruction quality using unseen poses with the interpolation capability aspect. Similarly, we sample every 4th frame in the CAPE dataset~\cite{ma2020cape} and measure the reconstruction accuracy on every second frame excluding the training frames. The results are shown in \Fig{Comparison[Meta]}.
In~\Tbl{CAPE}, we show quantitative results, and the scores of our approach rate better than MetaAvatar.

\begin{figure}[t]
	    \includegraphics[width=0.46\textwidth]{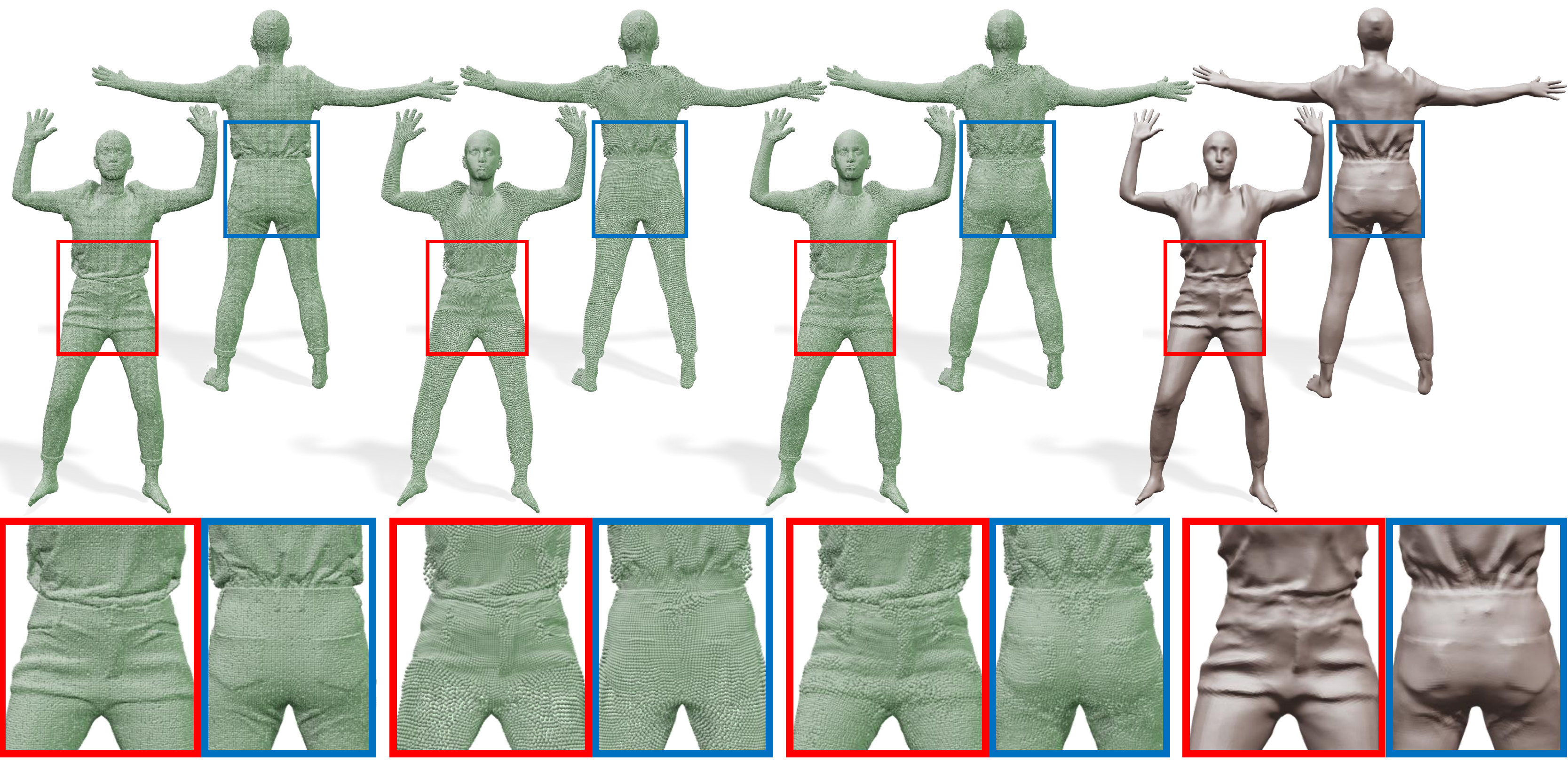}\\
	    \includegraphics[width=0.46\textwidth]{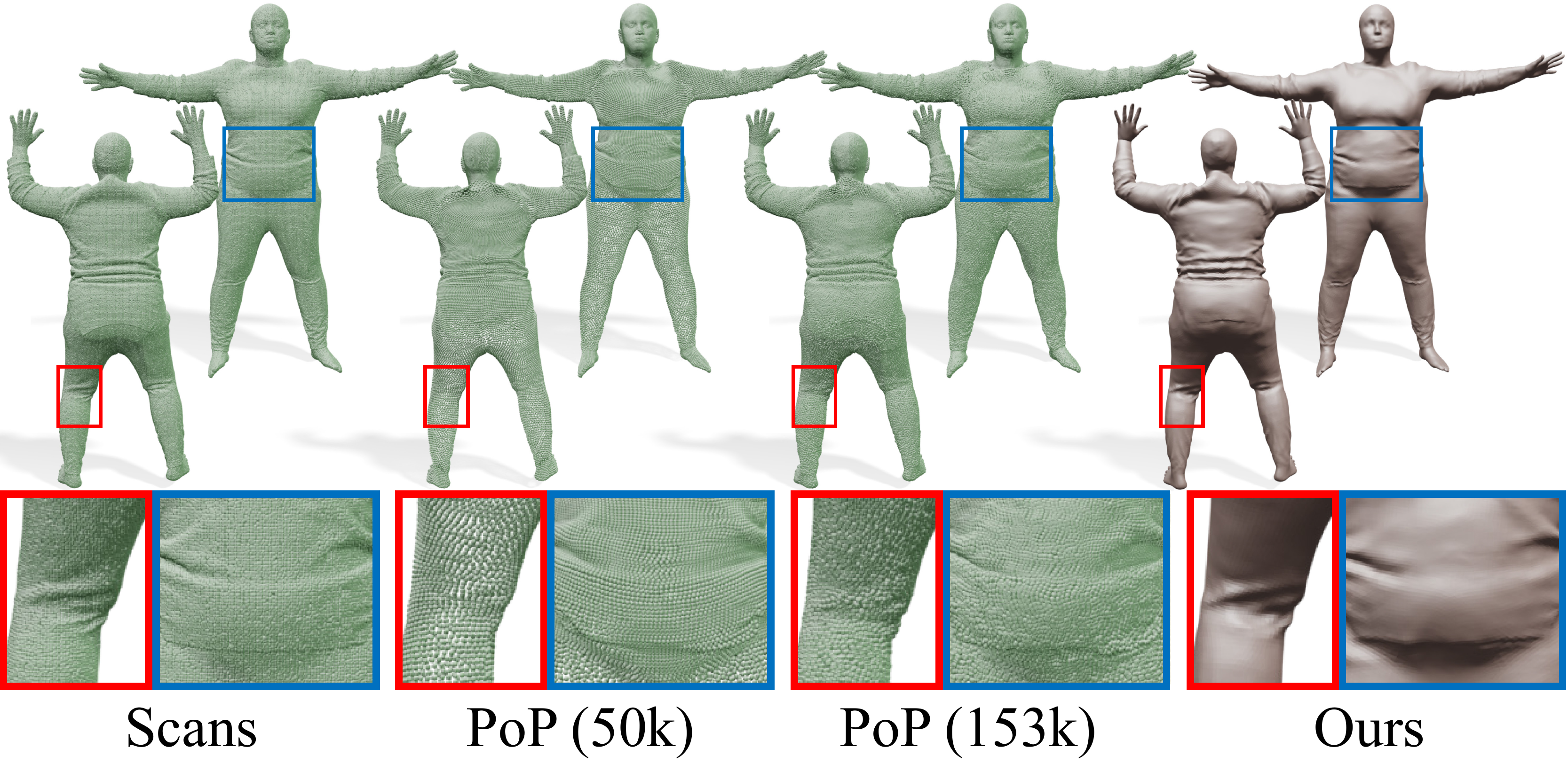}\\
	\vspace{-2mm}
	\caption{
	Comparison with two different settings of PoP~\cite{POP:ICCV:2021} on Resynth~\cite{POP:ICCV:2021} dataset. (From left to right) raw 3D point clouds, results of PoP (coarse and dense), and our results. Although our mesh has 113k vertices, it can sufficiently preserve geometric details with the aid of Laplacian coordinates.
	}
	\label{fig:Comparison[POP]}
\end{figure}

\begin{table}[t]
    \centering
    \caption{Comparisons with MetaAvatar \& PoP on CAPE~\cite{ma2020cape}.} 
    \begin{tabular}{c| c| c}
        \hline
        Method & $d_{CD}$(cm) $\downarrow$ & $NC$ $\uparrow$ \\
         \hline
         MetaAvatar~\shortcite{wang2021metaavatar} & 0.47 & 0.946\\
         \hline
         PoP~\shortcite{POP:ICCV:2021} (50k) & \textbf{0.32} & \textbf{0.977} \\
         \hline
         Ours & 0.45 & 0.959 \\
         \hline
    \end{tabular}
	\vspace{-2mm}
    \label{tbl:CAPE}
\end{table}

\paragraph{Comparison with PoP (Point cloud)}
Ma et al.~\shortcite{POP:ICCV:2021} utilize point cloud representation to deal with topological changes efficiently. We evaluate their method on CAPE dataset in the same configuration used for MetaAvatar in \Tbl{CAPE}. 
In addition, we compare our results with PoP on Resynth dataset~\cite{POP:ICCV:2021}, which has various cloth types with abundant details. However, Resynth dataset includes human models wearing skirts that cannot be properly handled by our framework (\Fig{Failure}), so we do not conduct quantitative evaluation using all subjects. Instead, we select five subjects not wearing skirts to compare our results with PoP quantitatively. We then split the training and test datasets in the same manner as on CAPE dataset. 

The original implementation of PoP queries the feature tensor with a $256\times 256$ UV map, resulting in a point cloud with 50k points. Since they are insufficient for representing details on the Resynth dataset, we modified the code to adopt $512\times 512$ UV map, and obtain 191k points. \Tbl{carla} shows quantitative comparisons with the above two settings, and our reconstruction is comparable with the original PoP. \Fig{Comparison[POP]} presents qualitative results where our mesh is more plausible than the original PoP, and comparable to dense PoP.

\begin{figure}[t]
	\centering
	    \includegraphics[width=0.47\textwidth]{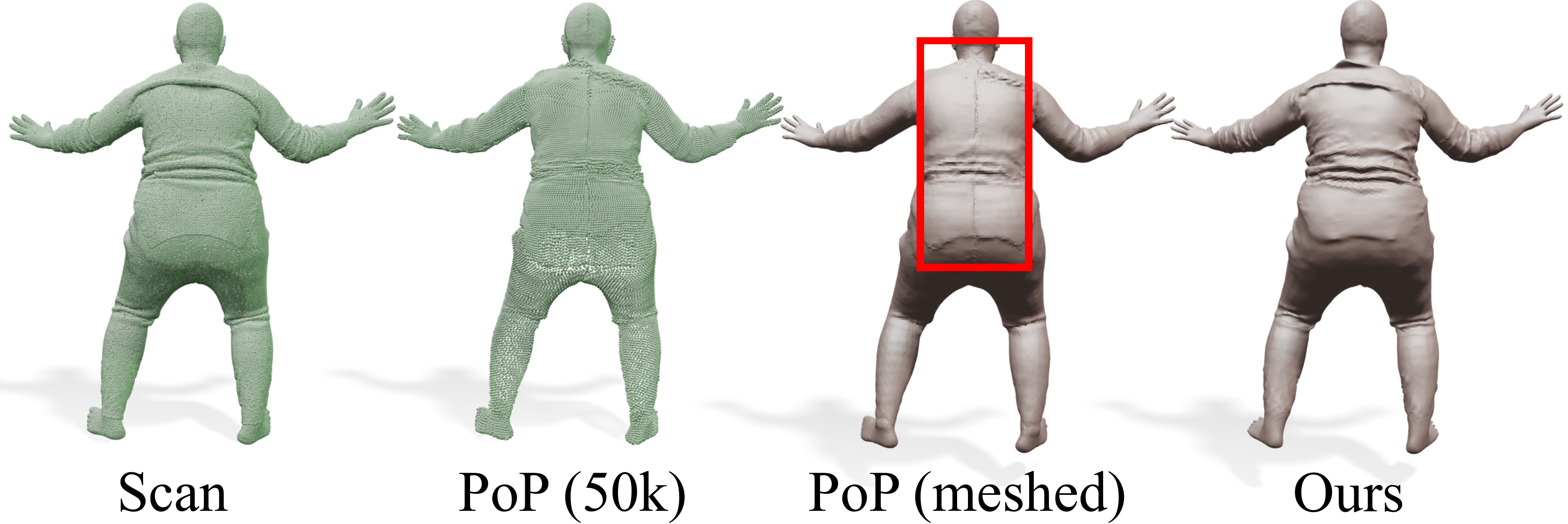}\\
	\vspace{-2mm}
	\caption{Seam artifacts of PoP~\cite{POP:ICCV:2021}. (From left to right) raw 3D point cloud, result of PoP, mesh result of PoP obtained using screened Poisson reconstruction~\cite{kazhdan2013screened}, and our result. In the red box, we highlight seam artifacts of PoP.
	}
	\label{fig:Comparison[POP_back]}
\end{figure}

In~\Fig{Comparison[POP_back]}, the mesh result obtained from a reconstructed point cloud of PoP contains a vertical line on the back because PoP uses UV coordinates for shape inference. In contrast, our approach can reconstruct clear wrinkles without such artifacts. 

\begin{table}[t]
    \centering
    \caption{Comparison with PoP on Resynth~\cite{POP:ICCV:2021}}
    \begin{tabular}{c| c| c| c}
        \hline
        \multicolumn{2}{c|}{Method} & $d_{CD}$(cm) $\downarrow$ & $NC$ $\uparrow$ \\
         \hline
         \multirow{2}{*}{PoP~\shortcite{POP:ICCV:2021}} & (50k)  & 0.43 & \textbf{0.970}\\
         \cline{2-4}
         & (153k)  & \textbf{0.33} & 0.968\\
         \hline
         \multicolumn{2}{c|}{Ours} & 0.41 & 0.964\\
         \hline
    \end{tabular}
	\vspace{-2mm}
    \label{tbl:carla}
\end{table}

\begin{figure}[t]
	\centering
	\includegraphics[width=0.47\textwidth]{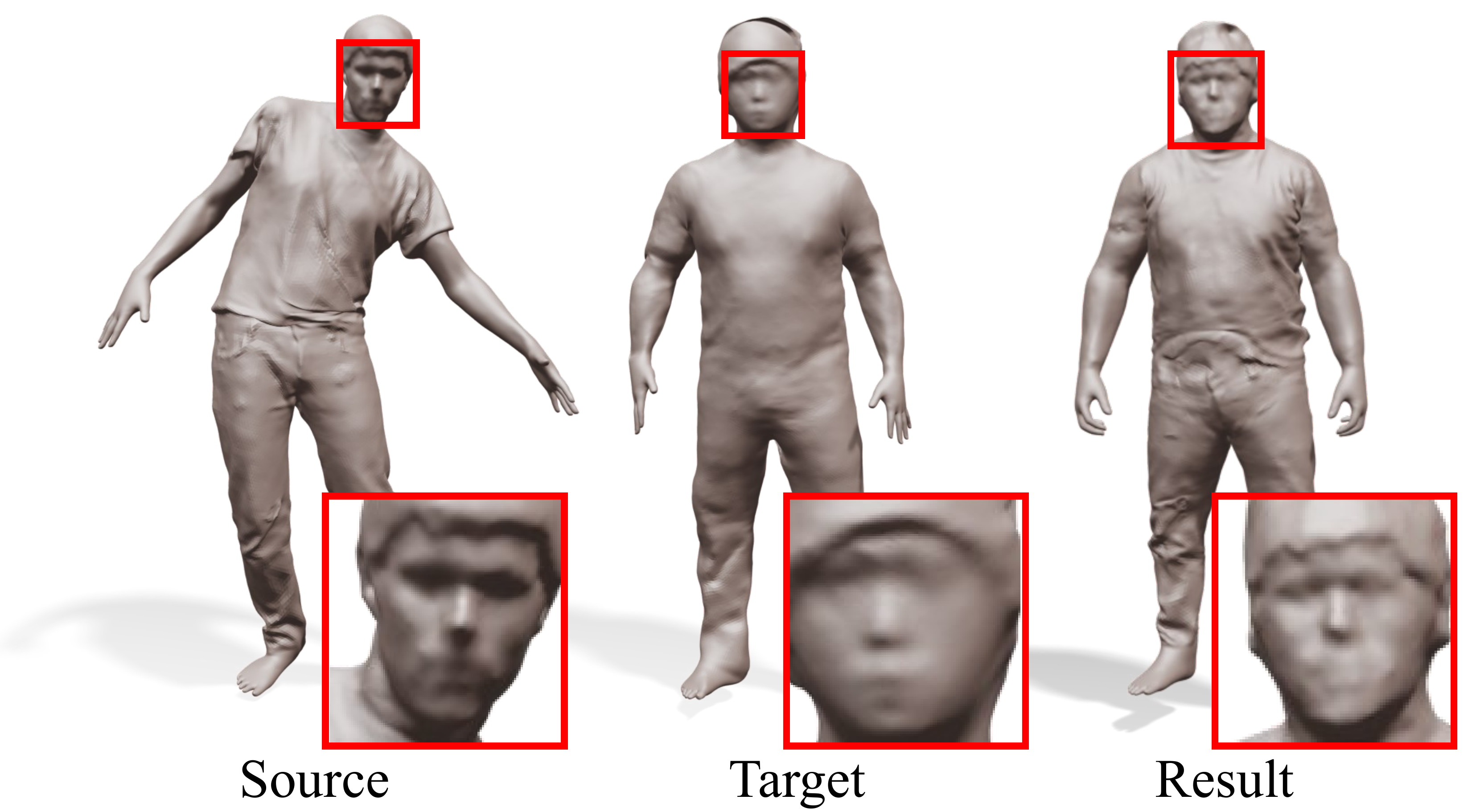} \\
	\includegraphics[width=0.47\textwidth]{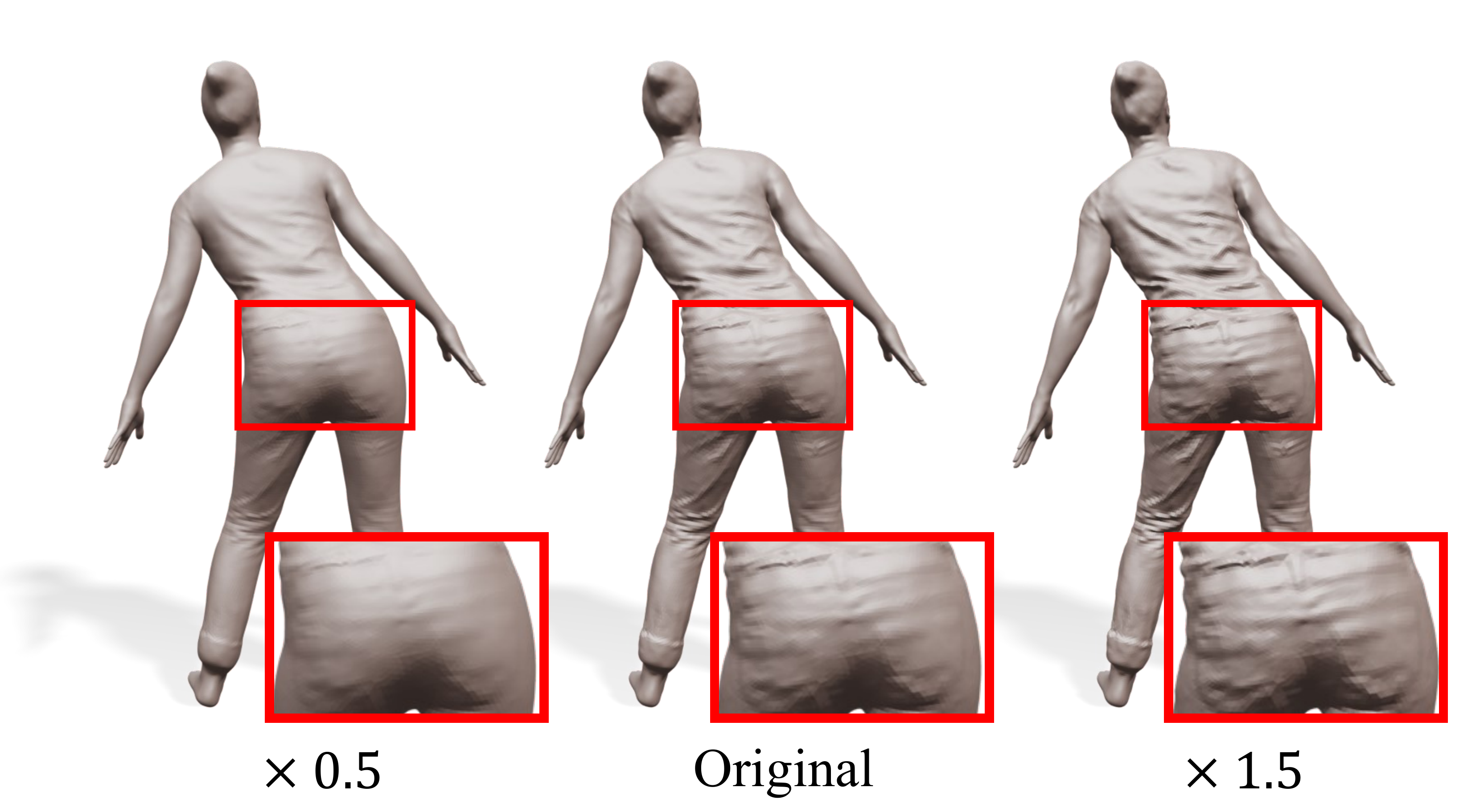}\\
	\vspace{-3mm}
	\caption{Detail transfer (top) and smoothing \& sharpening (bottom). (top) In our method, the optimized neural surface Laplacian function can be applied to another subject. (bottom) The amount of details can be easily adjusted by scaling Laplacian coordinates.}
	\label{fig:Application[enhancetransfer]}
	\vspace{-3mm}
\end{figure}

\subsection{Applications}
\label{sec:application}
\paragraph{Detail transfer}
Our neural surface Laplacian function encodes detailed shape information as the Laplacian coordinates defined at query points in a common domain. As a result, we can easily transfer shape details to other models by evaluating the original surface Laplacian function on the target pose-dependent base mesh. \Fig{Application[enhancetransfer]} shows an example.

\paragraph{Sharpening \& Smoothing}
Laplacian coordinates predicted by a neural surface Laplacian function in the Laplacian reconstruction step (\Sec{reconstruction}) can be scaled to change the amount of reconstructed shape details.
Multiplying a value greater than 1.0 to the predicted Laplacian coordinates performs detail sharpening, and the opposite performs detail smoothing. \Fig{Application[enhancetransfer]} shows examples.

\begin{figure}[t]
	\includegraphics[width=0.47\textwidth]{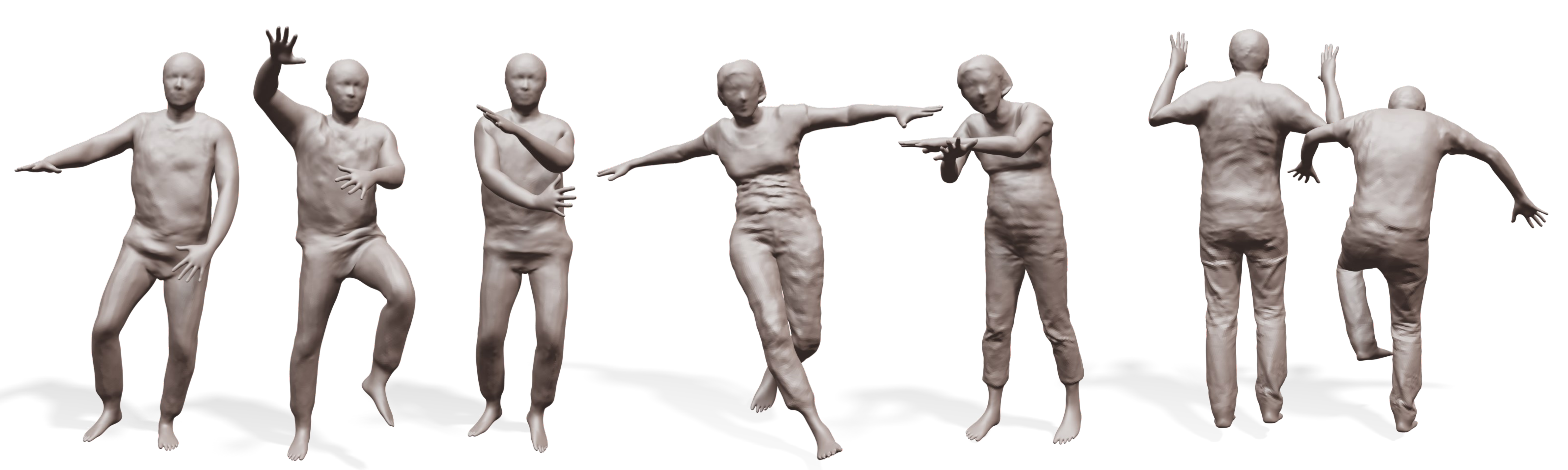} \\
	\vspace{-2mm}
	\caption{Animating examples. See our supplementary video.}
	\vspace{-3mm}
	\label{fig:Application[motion]}
\end{figure}

\paragraph{Animating}
\label{sec:animate}
The pose parameters used for evaluating surface functions $f_d$ and $f_l$ can be arbitrary.
In the case of reconstructing a scanned animation sequence, we would use the pose parameters $\boldsymbol{\theta}_t$ estimated from the input scans (\Sec{basemesh}).
On the other hand, once the functions $f_d$ and $f_l$ have been optimized, any parameters $\boldsymbol{\theta}$ other than $\boldsymbol{\theta}_t$ can be used to produce unseen poses of the subject.
In that case, the validity of shape details of the unseen poses is not guaranteed but our experiments generate reasonable results.
In \Fig{Application[motion]}, we optimized the functions $f_d$ and $f_l$ on BUFF and CAPE datasets and adopted synthetic motions from AIST++~\cite{li2021learn}. 
The results show natural variations of shape details depending on the motions.

\begin{figure}[t]
		\includegraphics[width=0.45\textwidth]{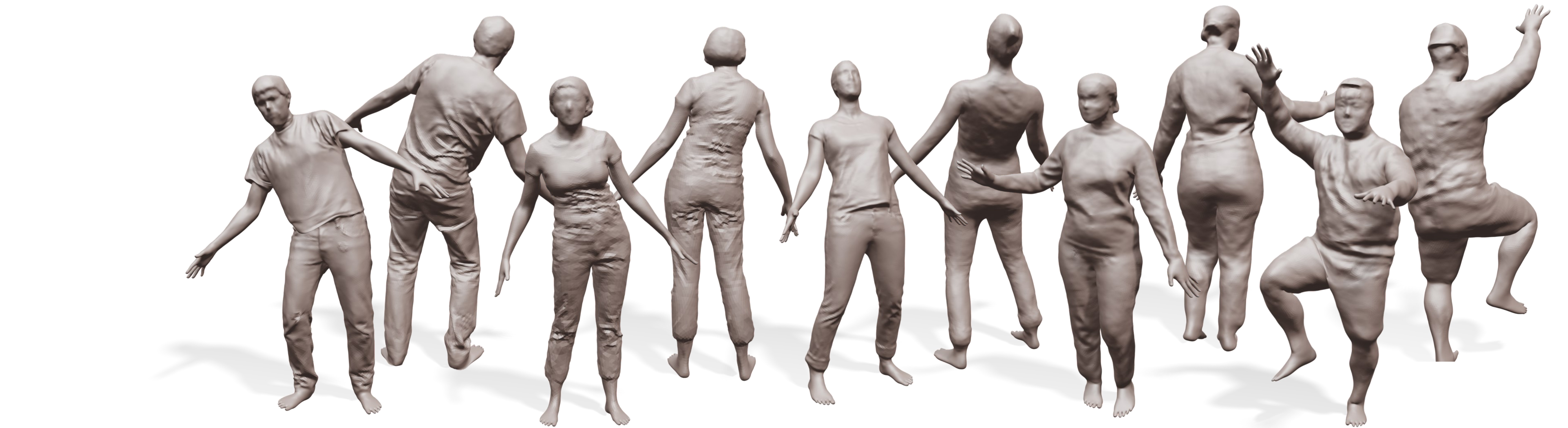} \\
		\includegraphics[width=0.45\textwidth]{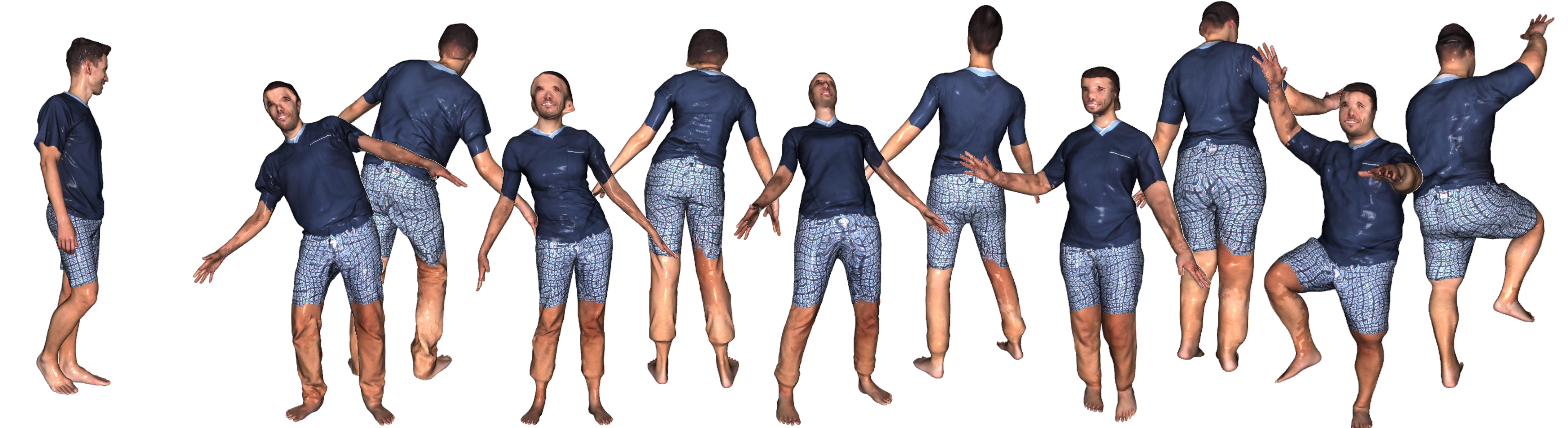} \\
		\includegraphics[width=0.45\textwidth]{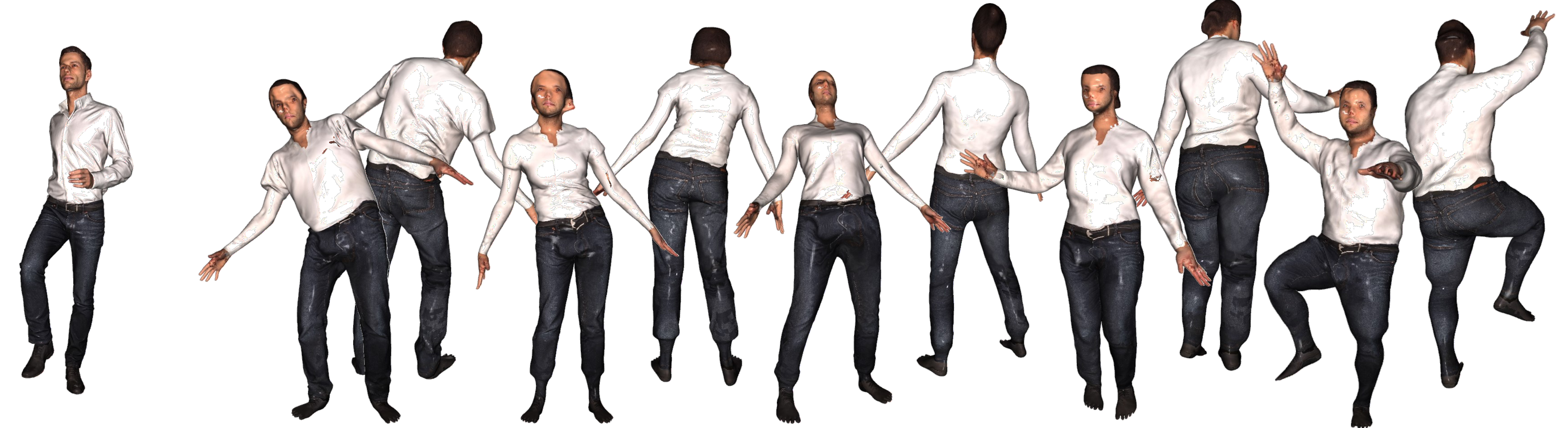} \\
		\includegraphics[width=0.45\textwidth]{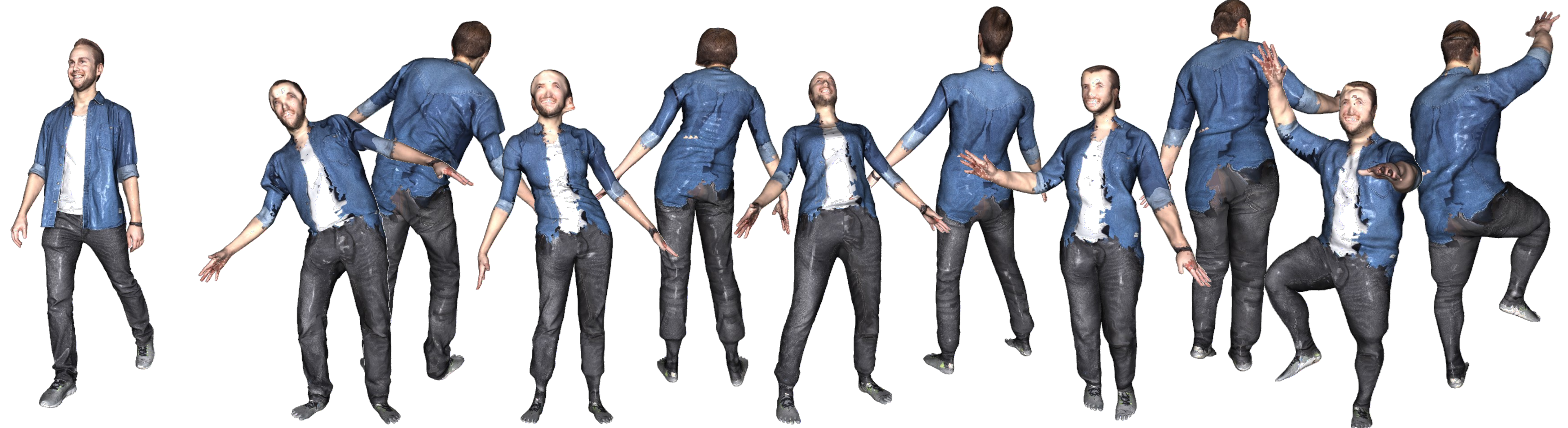} \\
		\includegraphics[width=0.45\textwidth]{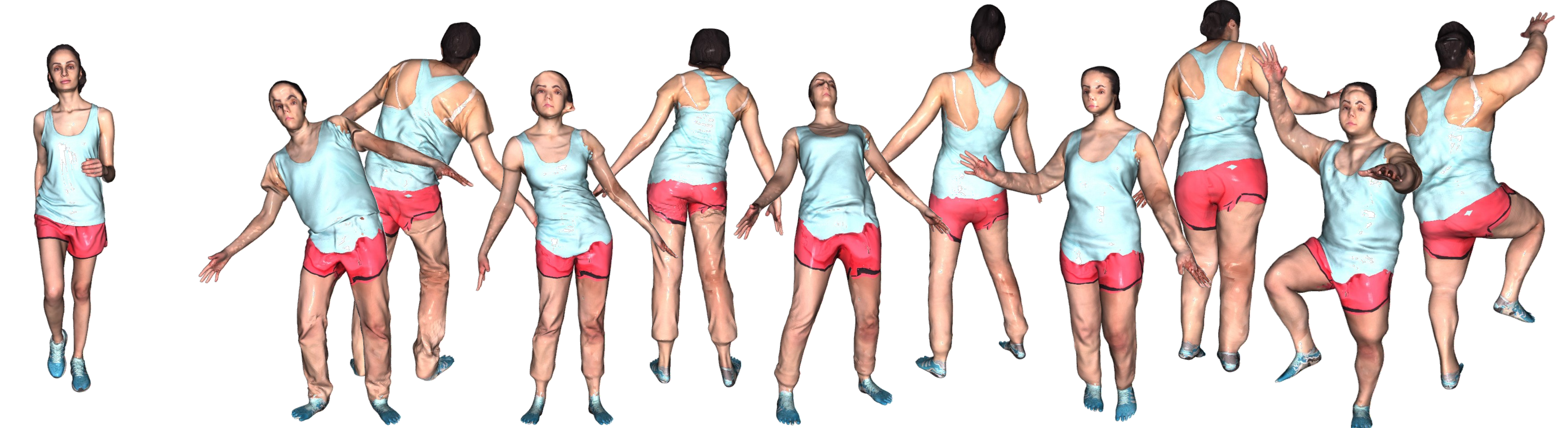} \\
		\includegraphics[width=0.45\textwidth]{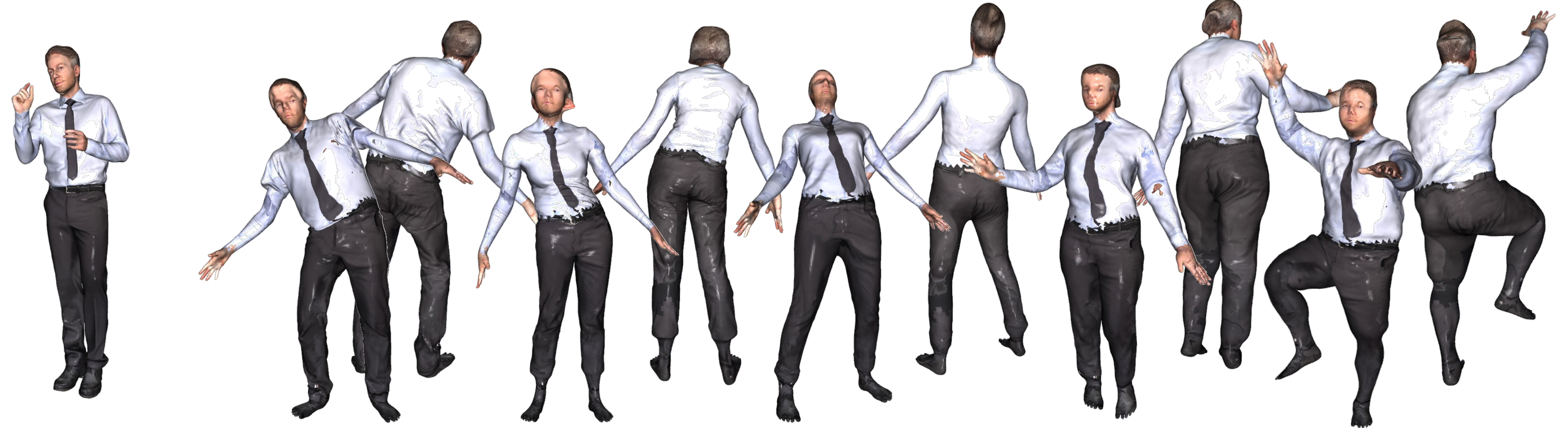} \\
		\includegraphics[width=0.45\textwidth]{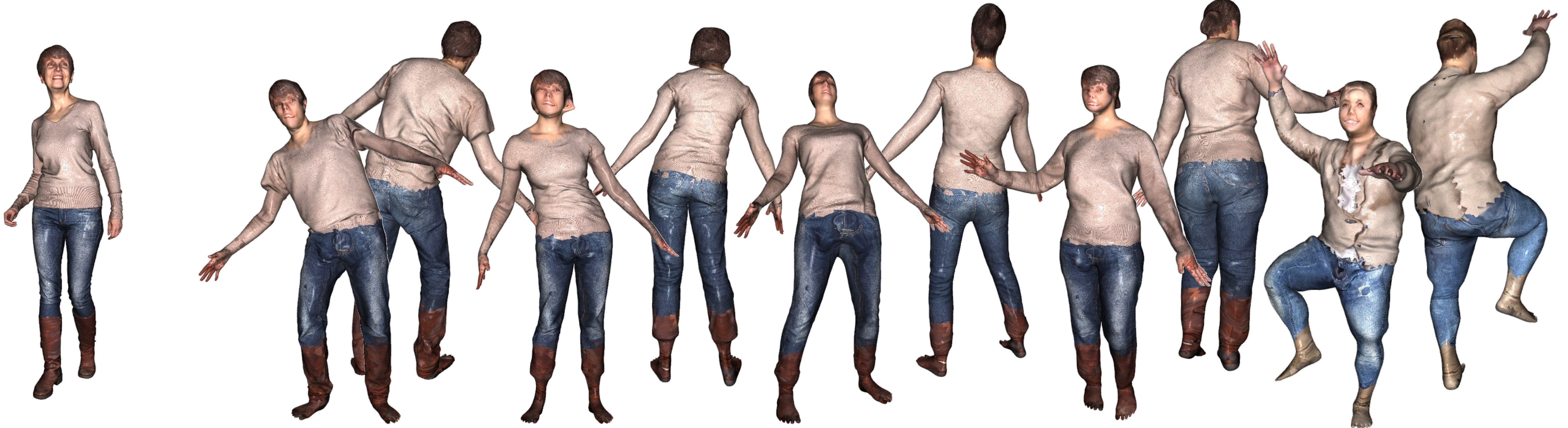} \\
	\vspace{-2mm}
	\caption{Texture mapping examples. (left) texture sources from RenderPoeple models~\cite{renderpeople}. (top) our reconstructed models. (others) texture transfer results. From the second column, each column shows the same shape with various textures. Since we use a fixed topology mesh, all our meshes have a common UV parametric domain. As a result, a single texture map can be shared among different reconstructed models without manual annotation.}
	\label{fig:Application[texture]}
	\vspace{-3mm}
\end{figure}

\paragraph{Texture mapping}
In our LaplacianFusion framework, all reconstructed models have the same fixed topology resulting from the subdivision applied to the SMPL model. 
Then, we can build a common UV parametric domain for texture mapping of any reconstructed model.
In \Fig{Application[texture]}, we initially transfer a texture from a RenderPoeple model~\cite{renderpeople} to the T-posed canonical neutral SMPL model $\mathcal{M}_C$ by using deep virtual markers~\cite{kim2021deep}. Then, the texture of $\mathcal{M}_C$ can be shared with different reconstructed models through the common UV parametric domain.

\section{Conclusions}
We presented a novel framework, \textit{LaplacianFusion}, that can reconstruct a detailed and controllable 3D clothed human body model from a 3D point cloud sequence. 
The key of our framework is Laplacian coordinates that can directly represent local shape variations.
We introduce a {\em neural surface Laplacian function} that uses Laplacian coordinates for encoding shape details from raw scans and then predicting desirable shape details on a pose-dependent base mesh.
The final model is reconstructed by integrating the Laplacian coordinates predicted on a subdivided base mesh.
Our approach can also be utilized for other applications, such as detail transfer. 

\begin{figure}[t]
	\includegraphics[width=0.35\textwidth]{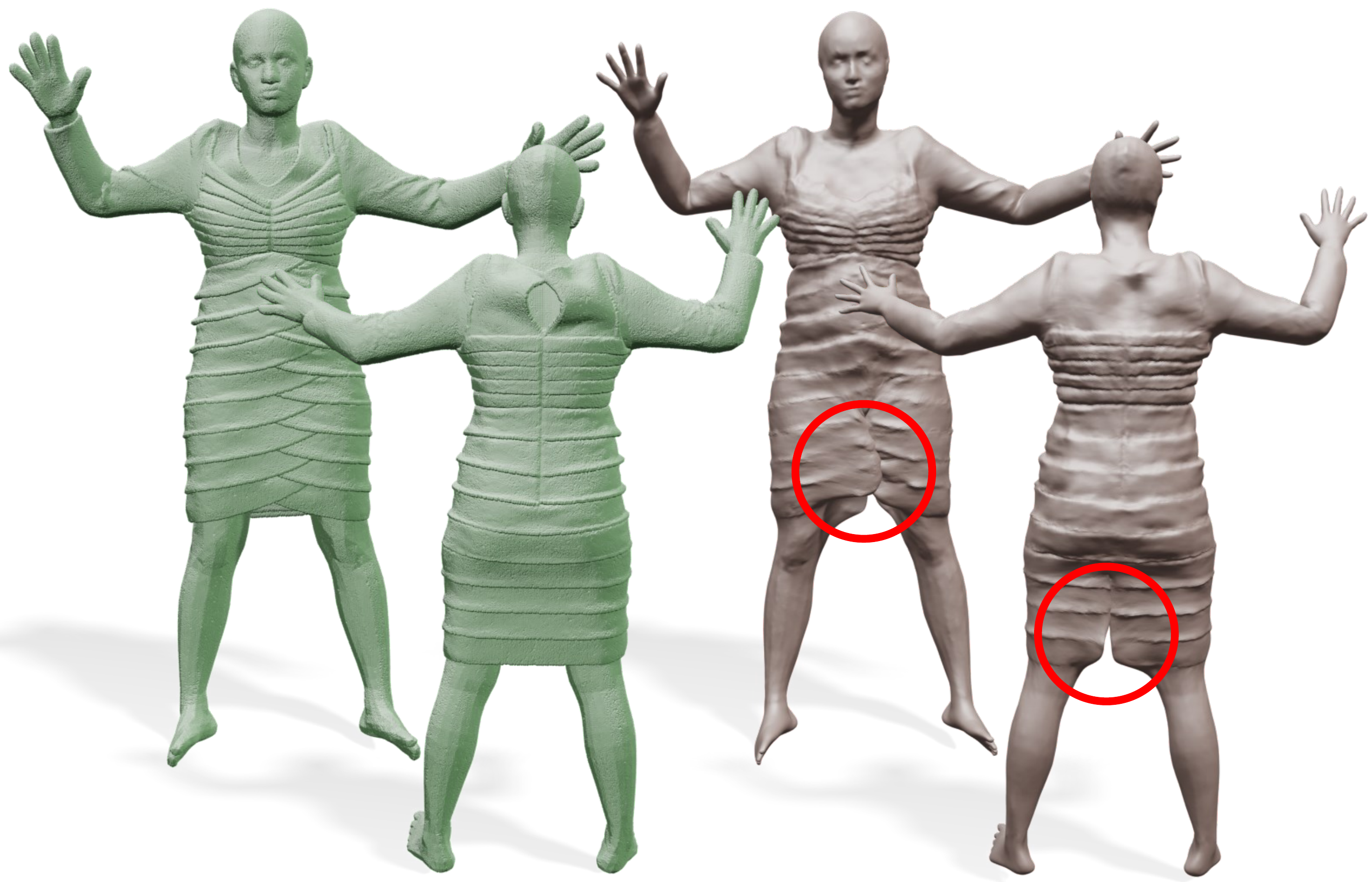} \\
	\vspace{-3mm}
	\caption{Failure case. (left) raw 3D point cloud of a human model wearing skirt in Resynth dataset~\cite{POP:ICCV:2021}. (right) our reconstruction result. Our pose-dependent base mesh cannot represent a seamless surface covering both legs (red circles) as its topology is originated from the T-pose of the SMPL model.}
	\vspace{-3mm}
	\label{fig:Failure}
\end{figure}

\paragraph{Limitations and future work} 
Since our framework uses a fixed topology mesh, we cannot cover topological changes, such as opening a zipper. 
In addition, our base mesh is initialized from a skinned body shape, so it is hard to deal with loose clothes, such as skirts (\Fig{Failure}).
Our framework relies on registration of the SMPL model to the input scan sequence, and the reconstruction quality is affected by the registration accuracy.
Currently, we use simple mid-point subdivision to increase the number of vertices in a base mesh, but a data-driven subdivision approach~\cite{liu2020neural} could be considered.
Our neural surface Laplacian function is trained for one subject, and generalization of the function to handle other subjects remains as future work.
We also plan to generalize our framework for non-human objects.


\begin{acks}
We would like to thank the anonymous reviewers for their constructive comments. This work was supported by IITP grants (SW
Star Lab, 2015-0-00174; AI Innovation Hub, 2021-0-02068; AI Graduate School Program (POSTECH), 2019-0-01906) and KOCCA grant
(R2021040136) from Korea government (MSIT and MCST).
\end{acks}

\newpage
\balance
\bibliographystyle{ACM-Reference-Format}
\bibliography{LaplFusion}

\end{document}